\documentclass[final,3p,times]{elsarticle}
\journal{Nuclear Physics B}
\pdfoutput=1
\usepackage{bm}
\usepackage{multirow}
\usepackage{cancel}
\usepackage{amsmath}
\usepackage{amssymb}
\usepackage{hyperref}
\biboptions{sort&compress}
\newcommand{\beq}{\begin{equation}}
\newcommand{\eeq}{\end{equation}}
\newcommand{\bea}{\begin{eqnarray}}
\newcommand{\eea}{\end{eqnarray}}
\newcommand{\bpmat}{\begin{pmatrix}}
\newcommand{\epmat}{\end{pmatrix}}

\begin{document}
\begin{frontmatter}{}
\title{\bf SU(5) unification of two triplet seesaw and leptogenesis with  dark matter and vacuum stability}
\author[1]{Mina Ketan Parida}\corref{cor1}%
\ead{minaparida@soa.ac.in}
\author[1]{Riyanka Samantaray}
\ead{riyankasamantaray59@gmail.com}
\address[1]{Centre of Excellence in Theoretical and Mathematical Sciences, Siksha 'O' Anusandhan, Deemed to be University, Khandagiri Square, Bhubaneswar 751030, India}   
\cortext[cor1]{Corresponding author}
\begin{abstract}
We investigate unification prospects of two heavy scalar triplet extension of the standard model where, in the absence of any right-handed neutrino (RHN), type-II seesaw  accounts for current 
 oscillation data with hierarchical neutrino masses consistent  with cosmological bounds and the lighter triplet decay explains 
baryon asymmetry of 
the Universe via leptogenesis. We note that  the absence of RHNs in the fundamental fermion representations of SU(5) delineates its outstanding position compared to SO(10) (or $E_6$). In addition, SU(5) needs smaller scalar representations ${15}_{H1}\oplus {15}_{H2}$  compared to  much larger representations ${126}_{H1}\oplus {126}_{H2} \subset $ SO(10) (or ${351}^{\prime}_{H1}\oplus {351}^{\prime}_{H2} \subset E_6$). We show how precision gauge coupling unification is achieved through SU(5) with the predictions of different sets of two heavy triplet masses which,  besides being compatible with type-II seesaw, are also consistent with unflavoured or $\tau -$ flavoured leptogenesis predictions for baryon asymmetry of the Universe. In addition to an intermediate mass colour octet fermion, completion of precision gauge coupling unification is found to require essentially the presence of the well known weak triplet fermion  $\Sigma (3,0,1)$ in its  mass
range  $M_{\Sigma}\simeq {\cal O}(500-3000)$ GeV  out of which the dominant dark matter (DM) resonance mass 
 $M_{\Sigma}\ge 2.4$ TeV is known to account for the observed  cosmological relic density. The deficiency in relic density for the other class of lighter $M_{\Sigma}$ solutions allowed under indirect search constraints 
 is circumvented by the introduction of a scalar singlet DM which could be as light as $62$ GeV. A GUT ansatz is noted to ensure  vacuum stability  of the SM scalar potential for all types of unification solutions realised in this work.  
 We discuss proton lifetime estimations for $p\to e^+\pi^0$ compatible with the present Hyper-Kamiokande bound as a function of an unknown mixing parameter in the model.
\end{abstract}
\end{frontmatter}{}
\section{Introduction}\label{sec.1}
Neutrino oscillation \cite{nudata,Forero:2014,Esteban:2018}, baryon asymmetry of the universe
\cite{BAUexpt,Planck15}, and  dark matter \cite{DMexpt,Planck15,wmap,Akerib:2016,Aprile:2017,Aprile:2018,Cui:2017,WIMP,GAMBIT} are
the three  prominent physics issues beyond the purview of the 
standard model (SM). Further, the origin of different strengths of strong, weak, and electromagnetic interactions in $SU(2)_L\times U(1)_Y\times SU(3)_C(\equiv G_{213})$ can not be explained by SM itself \cite{JCP:1974,su5,so10,E6,cmp:1984,mkp-rs:2020}. However, the underlying origin of neutrino masses is widely recognised to be seesaw mechanisms \cite{type-I,Valle:1980,type-II,type-III} some of which also explain baryon asymmetry of the Universe (BAU) via leptogenesis caused by the decay of mediating heavy scalars or fermions, and by sphaleron interactions \cite{sphaleron}.   
Large number of leptogenesis models using right-handed neutrino (RHN) mediated type-I   \cite{type-I,Valle:1980,Fuku-Yana:1986} and other seesaw mechanisms have been  proposed for successful baryon asymmetry generation;  a partial list of such extensive investigations has been given in \cite{Nir:2008,psna:2017,cps:2019,ASJWR} and references therein.
In a novel interesting proposal  \cite{Ma-Us:1998} as an alternative mechanism  without  RHNs, realization
of neutrino masses and baryon asymmetry of the Universe (BAU) has been  shown to be
possible \cite{Joaquim:2013,Sierra:2014tqa,Sierra:2011ab,pcns:2020} through the SM extension  by two heavy scalar triplets, $\Delta_1(3,-1,1)$ and $\Delta_2 (3,-1,1)$ of identical $G_{213}$ charges, each of which also generates neutrino mass by  type-II seesaw \cite{type-II}. 
The tree level dilepton decay of any one of these
triplets combined with loop contribution generated by their
collaboration predicts the desired CP-asymmetry
formula for leptogenesis leading to observed baryon asymmetry of the Universe.\\  

 Very recently, as a  departure from  the proposed applicability to quasi-degenerate neutrinos \cite{Ma-Us:1998},  we have shown \cite{pcns:2020} the success of the two triplet model (TTM) in following respects:  The model explains current neutrino data \cite{nudata,Forero:2014,Esteban:2018} comprising of atmospheric neutrino mixing in the second octant, large Dirac CP phases, and hierarchical neutrino masses of both normal and inverted orderings. The model has been also found to be consistent with the  cosmological bound $\Sigma_{(2015)}$ on the sum of three neutrino masses  derived using the Planck satellite data  \cite{Planck15} as well as the improved  bound $\Sigma_{(2018)}$ \cite{Sunny:2018,Planck18-1,Planck18-2}
\beq
\Sigma_{(2015)}=\sum_{i=1}^3 {\hat m}_i \le 0.23\,{\rm eV};\,\, \Sigma_{(2018)}=\sum_{i=1}^3 {\hat m}_i \le 0.12\, {\rm eV}. \label{eq:sig1518}
\eeq
Additional sources of CP-asymmetry of TTM have been exploited to  further  predict the BAU in the cases of unflavoured and partially flavoured ($\equiv \tau-$
flavoured) leptogenesis covering lighter triplet mass $M_{\Delta_2}\simeq {\cal O}(10^{10}-10^{14})$ GeV and the corresponding values of heavier triplet masses $M_{\Delta_1}\ge 10 M_{\Delta_2}$ \cite{pcns:2020}. Moreover, the two triplet model (TTM) has been shown to possess an inbuilt
fine tuning  mechanism to keep the standard Higgs mass safe against radiative instability \cite{pcns:2020}. With additional $Z_2$ stabilising discrete symmetry, the model has also successfully addressed issues on dark matter (DM) and vacuum sability with the prediction of a scalar DM mass  $\sim 1.3$ TeV which also ensures 
radiative stability of the standard Higgs  scalar mass \cite{pcns:2020}.

In this work, for the first time, we implement unification of three gauge couplings of the two triplet model (TTM) which has been left unanswered so far.  New origins of CP-asymmetry in hybrid seesaw \cite{psna:2017} and in type-II seesaw dominated single-triplet leptogenesis model empowered by RHN loop mediation have been successfully addressed  in SO(10) \cite{mkp-rs:2020,cps:2019}. Triplet seesaw dominance has been shown to predict $\nu_\mu-\nu_\tau$ mixing in SUSY GUTs \cite{Bajc-gs-Vissani:2003,Goh-RNM-Ng:2003} as well as three neutrino mixings \cite{RNM-mkp-gr:2004} and fermion masses in SO(10) \cite{ASJ-KMP:2011,mkp:2008}. Also there are  interesting investigations in left-right symmetric models or in SM extensions with or without dark matter \cite{Ham-gs:2003,Ham-Ma:2006,Ham-Strum:2006,Ham:2012,Gu-Ma-Us:2016,Rink:2020,Majee:2007,Ohlsson:2020,Wetterich:2021}.\\
As SO(10) \cite{so10} and $E_6$ \cite{E6,Slansky:1979} fundamental fermion representations intrinsically contain  RHNs, in this work we search for an alternative GUT since the  TTM is based upon the unconventional novel hypothesis that the origin of neutrino masses
as well as BAU generation do not need RHNs. However,  the SU(5) theory \cite{su5,Nath-Perez:2007,Perez:2019-1,Perez:2018-2} does not have
RHNs in its fundamental fermion representations: ${10}_F\oplus {\bar 5}_F$ which exactly complete the SM fermions. We would like to emphasize that, unlike the SU(5) models occurring as intermediate symmetry  in SO(10)(or $E_6$) $\to$ SU(5) $\to$ SM \cite{cps:2019,Goh-RNM-Nasri:2004,RNM-mkp:2011}, here we utilise the independent class of  SU(5) models \cite{su5,Nath-Perez:2007,Perez:2019-1,Perez:2018-2}. Hybrid seesaw near TeV scale mediated by triplet and singlet fermions
has been implemented proposing interesting experimental signature at LHC \cite{Bajc-Nem-gs:2007}. 
Noting further that a scalar triplet $\Delta(3,-1,1)$
is contained in the representation  ${15}_{H}$ of SU(5), or in  ${126}_{H}$ of SO(10)
(or in ${351}^{\prime}_{H}$ of $E_6$), the embedding of the two scalar triplets requires much smaller SU(5) scalar representations ${15}_{H1}\oplus {15}_{H2}$ compared to far more larger representations ${126}_{H1}\oplus {126}_{H2} \subset  SO(10)$ and ${351}^{\prime}_{H1}\oplus {351}^{\prime}_{H2} \subset E_6$.  
We show how such a SU(5) GUT framework \cite{su5} predicts 
all the ingredients of the TTM \cite{pcns:2020}  with gauge unification of strong, weak, and electromagnetic interactions. Interestingly, the completion of coupling unification is found to predict the well known non-standard fermion triplet $\Sigma(3,0,1)$ as an important component of WIMP dark matter  \cite{Cirelli:2006,Cirelli:2007,Ma-Suematsu:2008,Cirelli:2008,Franceschini:2008,Frig-Ham:2010,psb:2010} where vacuum stability of the SM Higgs scalar potential is ensured by an intermediate mass scalar singlet threshold effect \cite{Espinosa:2012}.  Proton lifetime formula in this class of SU(5) contains an unknown mixing parameter \cite{Perez:2019-1,Perez:2018-2} whose allowed range in the present TTM unification realization is only partially accessed by the  current  Super-Kamiokande \cite{SuperK} or Hyper-Kamiokande search limits \cite{HyperK}.
Highlights of the present work are\\ 
\begin{itemize}
\item{The absence of RHN in the fundamental fermion representation of  SU(5)
compared to SO(10) and $E_6$ is noted to identify the minimal rank-4 GUT as a natural candidate to unify the two-triplet model (TTM). In addition, this  SU(5) embedding of TTM requires 
smaller scalar representations ${15}_{H1}\oplus {15}_{H2}$  compared to much larger  representations ${126}_{H1}\oplus {126}_{H2}$ of SO(10), or ${351}^{\prime}_{H1}\oplus {351}_{H2}^{\prime}$ of $E_6$.} 
\item{Renormalisation group (RG) predicted successful embedding of the two triplet model in SU(5) grand unified theory with  precision gauge coupling unification at $M_U\simeq {\cal O}(10^{15})$ GeV. }
\item{Unification prediction of all the sets of two heavy scalar triplet masses \cite{pcns:2020} in the range 
${\cal O} (10^{10})\,{\rm GeV} < M_{\Delta_2} <  M_{\Delta_1}\le M_U$ ensuring type-II seesaw ansatz for neutrino oscillation data fitting and baryon asymmetry prediction through unflavoured or $\tau-$ flavoured leptogenesis \cite{pcns:2020}.} 
\item{ In addition to an intermediate mass colour octet fermion, the  unification completion is found to predict essentially a fermionic triplet dark matter $\Sigma(3,0,1)$  \cite{Cirelli:2006,Ma-Suematsu:2008,Frig-Ham:2010,psb:2010} in its mass range $M_{\Sigma}\simeq {\cal O}(500-3000)$ GeV out of which dominant minimal thermal DM  with resonance mass $M_{\Sigma}= 2.4$ TeV, or heavier,  accounts for the entire observed value of cosmological relic density.}
\item{In the other class of lighter fermionic triplet sub-dominant DM mass solutions, $M_{\Sigma}\simeq {\cal O}(500-2000)$ GeV, permitted by indirect search constraints, the model is shown to account for the observed DM relic density through the introduction of  a scalar singlet DM $\xi$ whose mass  could be as light as $m_{\xi}\simeq 62$ GeV.}  
\item{  An intermediate mass Higgs scalar singlet threshold effect originating from SU(5) breaking is noted to ensure vacuum stability of the SM scalar potential for all classes of
allowed solutions emerging from this GUT realization.}
\item{ In view of an unknown mixing parameter  that occurs in the proton decay formula for $p\to e^+\pi^0$ in this class of  SU(5) models \cite{Perez:2019-1,Perez:2018-2}, the  present  Hyper-Kamiokande limit \cite{HyperK} is capable of constraining only a limited region of this parameter space.}  
\end{itemize}
This paper is organised in the following manner.
 In Sec.\ref{sec:ttm} we briefly summarise  the recent work on two triplet model without grand unification \cite{pcns:2020}.
Sec. \ref{sec:su5} elucidates our new implementation and unification results in SU(5) GUT framework including $\Delta_1, \Delta_2$ mass predictions required for neutrino mass fits and unflavoured or $\tau-$ flavoured leptogenesis. In  Sec.\ref{sec:taup},  we discuss  proton lifetime estimations in this SU(5) model. Sec.\ref{sec:fdm} deals with  $SU(5)\times Z_2$ ansatz for fermionic triplet dark matter (DM) and its prospects and limitations. Vacuum stability of the SM scalar potential is discussed in Sec.\ref{sec:vstab}.
The model capability to accommodate fermionic triplet plus scalar singlet DM is discussed in Sec.\ref{sec:fxidm}. Grand unification advantage over ununified TTM in predicting fermionic triplet or scalar singlet DM  and vacuum stability are briefly discussed in Sec. \ref{sec:GA}. We summarize and conclude  in Sec.\ref{sec:sum}. In Sec.\ref{sec:A1} of Appendix a brief discussion is made on non-standard fermion masses.  
\section{The two triplet model for neutrino mass and leptogenesis}\label{sec:ttm}
In this section we briefly summarise the outcome of the two-triplet model \cite{pcns:2020} relevant for the present embedding in  SU(5) grand unification.
\subsection{\textbf{Neutrino masses and mixings through type-II seesaw}}\label{sec:numass}
The SM is extended by the addition of two heavy scalar triplets $\Delta_1 (3,-1,1)$ and $\Delta_2(3,-1,1)$ with masses $M_{\Delta_1}$ and $M_{\Delta_2}$, respectively. Here respective charges have been indicated under the SM gauge group $SU(2)_L\times U(1)_Y \times SU(3)_C$. The extended 
part of the Lagrangian that contributes to type-II seesaw generation of neutrino masses and leptogenesis is
\bea
-{\cal L}_{ext} &=&\sum_{\alpha=1}^2\left ((D_{\mu}{\vec {\Delta}}_{\alpha})^{\dagger}.(D^{\mu}{\vec {\Delta}}_{\alpha})- M^2_{\Delta_\alpha}
Tr(\Delta_\alpha^\dagger \Delta_\alpha) \right )\nonumber\\
&&+ \sum_{\alpha=1}^2\left ( [\frac{1}{2}\sum_{ij}f^{(\alpha)}_{ij} L^T_i Ci\tau_2 \Delta_\alpha L_j 
 - \mu_{\Delta_\alpha} \phi^T i\tau_2 \Delta_\alpha \phi
+h.c.]\right). \label{Yukhiggs}
\eea
 Here  the three lepton
doublets are $L_i (i=1,2,3)$, $\alpha=1,2$ denote  the two scalar triplets, $M_{\Delta_\alpha}=$  mass of the triplet $\Delta_\alpha$,  $f^{(\alpha)}_{ij}=$ 
Majorana coupling of $\Delta_\alpha$ with $L_i$ and $L_j$, and  $\mu_{\Delta_\alpha}=$  lepton-number violating (LNV) trilinear coupling of $\Delta_\alpha$ with
standard Higgs doublet $\phi$.    
The induced VEVs, $V_{\Delta_{\alpha}} (\alpha=1,2)$, parameterizing  neutrino mass matrix $m_{\nu}$ are
\beq
V_{\Delta_{\alpha}}= \frac{\mu_{\Delta_\alpha} v^2}{2 M_{\Delta_\alpha}^2}, (\alpha=1,2), \label{vlk} 
\eeq
\bea
m_{\nu} &=& 2 f^{(1)} V_{\Delta_{1}}+2 f^{(2)} V_{\Delta_{2}}, \nonumber\\
        &=& f^{(1)}\frac{\mu_{\Delta_1} v^2}{ M_{\Delta_1}^2}  +  f^{(2)} \frac{\mu_{\Delta_2} v^2}{ M_{\Delta_2}^2} \nonumber\\
&\equiv& m^{(1)}_{\nu}+m^{(2)}_{\nu}. \label{mnu}
\eea    
Here $v=246$ GeV $=$ the standard Higgs vacuum expectation value (VEV).
 The  neutrino data
are fitted 
using the PDG convention \cite{Beringer:2012} on the
PMNS mixing matrix
\begin{equation}
 U_{\rm{PMNS}}= \left( \begin{array}{ccc} c_{12} c_{13}&
                      s_{12} c_{13}&
                      s_{13} e^{-i\delta}\cr
-s_{12} c_{23}-c_{12} s_{23} s_{13} e^{i\delta}& c_{12} c_{23}-
s_{12} s_{23} s_{13} e^{i\delta}&
s_{23} c_{13}\cr
s_{12} s_{23} -c_{12} c_{23} s_{13} e^{i\delta}&
-c_{12} s_{23} -s_{12} c_{23} s_{13} e^{i\delta}&
c_{23} c_{13}\cr
\end{array}\right) 
diag(e^{\frac{i \alpha_M}{2}},e^{\frac{i \beta_M}{2}},1),
\end{equation}
where $s_{ij}=\sin \theta_{ij}, c_{ij}=\cos \theta_{ij}$ with
$(i,j=1,2,3)$, $\delta$ is the Dirac CP phase and $(\alpha_M,\beta_M)$
are Majorana phases. The  best fit values of the oscillation
data \cite{Forero:2014,Esteban:2018} summarised below in
Table \ref{tab:nudata}  have been used in \cite{pcns:2020}.

\begin{table}[!h]
\caption{Input data from neutrino oscillation experiments \label{osc} \cite{Forero:2014,Esteban:2018}}
\label{tab:nudata}
\begin{center}
\begin{tabular}{|c|c|c|}
\hline
{ Quantity} & {best fit values} &{ $3\sigma$ ranges}\\
\hline
$\Delta m_{21}^2~[10^{-5}eV^2]$ & $7.39$ & $6.79-8.01$\\
$|\Delta m_{31}^2|~[10^{-3}eV^2](NO)$ & $2.52$ & $2.427-2.625$\\
$|\Delta m_{32}^2|~[10^{-3}eV^2](IO)$ & $2.51$ & $2.412-2.611$\\
$\theta_{12}/^\circ$ & $33.82$ & $31.61-36.27$\\
$\theta_{23}/^\circ (NO)$ & $49.6$ & $40.3-52.4$\\
$\theta_{23}/^\circ (IO)$ & $49.8$ & $40.6-52.5$\\
$\theta_{13}/^\circ (NO)$ & $8.61$ & $8.22-8.99$\\
$\theta_{13}/^\circ (IO)$ & $8.65$ & $8.27-9.03$\\
$\delta/^\circ (NO)$ & $215$ & $125-392$\\
$\delta/^\circ (IO)$ & $284$ & $196-360$ \\
\hline
\end{tabular}
\end{center}
\end{table}
Important  features of the data 
are: (i)  atmospheric  mixing angle $\theta_{23}$  in the second octant, (ii) large values of Dirac CP phases exceeding $\delta=200^{\circ}$, (iii) includes the reactor neutrino mixing  $\theta_{13}=8.6^\circ$ determined earlier.\\
Denoting the mass eigen values as ${\hat m}_i(i=1,2,3)$ and the relation
 \begin{equation}
m_\nu = U_{PMNS}~diag({\hat m}_1, {\hat m}_2, {\hat m}_3) U_{PMNS}^T,\label{mnu}
\end{equation}
all the nine elements of the symmetric matrix $m_{\nu}$ for NO and IO type hierarchies including mass eigen values are given in Table \ref{tab:mnutab}.
\begin{table}[!h]
\caption{Neutrino mass matrices and eigen values in NO and IO hierarchies from oscillation data \label{matrix}.}
\label{tab:mnutab}
\begin{center}
\begin{tabular}{|c|c|}
\hline
{ Quantity} &  Best fit values\\
\hline
$(m_{\nu})_{ij} (\rm eV)(\rm NO)$&$\left( \begin{array}{ccc}
              0.00367-0.00105i  & -0.00205+0.00346i & -0.00634+0.00294i  \cr
               -0.00205+0.00346i & 0.03154+0.00034i & 0.02106-0.0001i \cr
                -0.00634+0.00294i & 0.02106-0.0001i & 0.02383-0.00027i\cr
                    \end{array}\right)$\\ 
\hline
$(m_{\nu})_{ij} (\rm eV) (\rm IO)$&$\left( \begin{array}{ccc}
                  0.0484-0.00001i & -0.001122+0.0055i & -0.00137+0.00471i \cr
                   -0.001122+0.0055i  &  0.02075-0.00025i& -0.02459-0.00026i \cr
                    -0.00137+0.00471i & -0.02459-0.00026i &  0.02910-0.00026i\cr
                    \end{array}\right)$\\
\hline
${\hat m}_i(\rm eV)\,(\rm NO)$&${\rm diag}(0.001,\,\, 0.0086,\,\, 0.0502)$\\
\hline
${\hat m}_i(\rm eV)\, (\rm IO)$&${\rm diag}(0.0493,\,\,0.0501,\,\,0.0010)$\\
\hline
$\sum_{(i=1)}^3{\hat m}_i(\rm eV)\,(\rm NO)$&$0.0598$\\
\hline 
$\sum_{(i=1)}^3{\hat m}_i(\rm eV)\,(\rm IO)$&$0.1000$\\
\hline 
\end{tabular}
\end{center}
\end{table}

The best fits in both the NO and IO cases satisfy the current cosmological bounds of eq.(\ref{eq:sig1518}) on the sum of three neutrino masses  which have been derived using the Planck satellite data and the $\Lambda$CDM \cite{Planck15,Sunny:2018} big-bang cosmology.

\subsection{\textbf{ CP-asymmetry}}

\begin{figure}
\begin{center}
 \includegraphics[width=4cm,height=3cm]{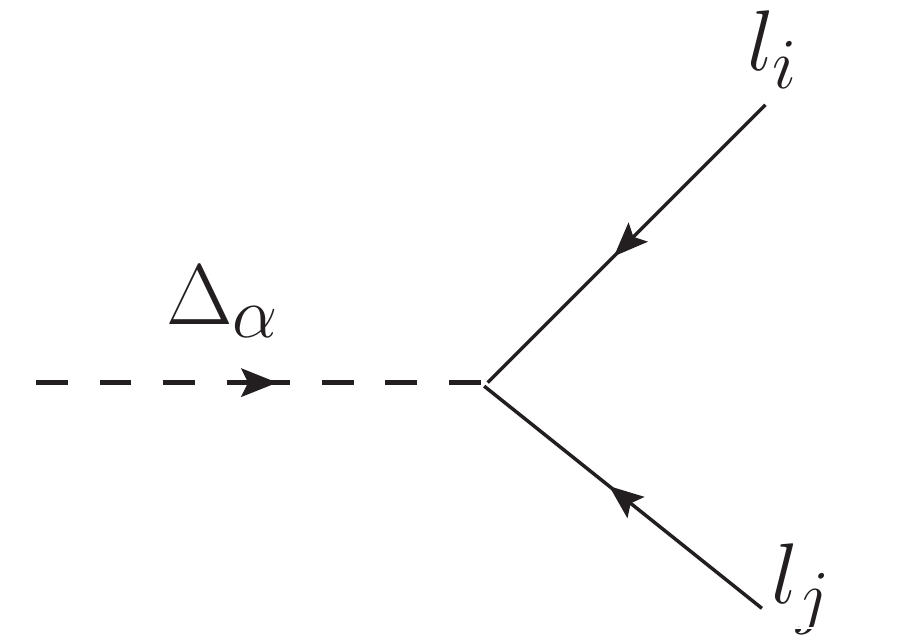}
 \includegraphics[width=5cm,height=3cm]{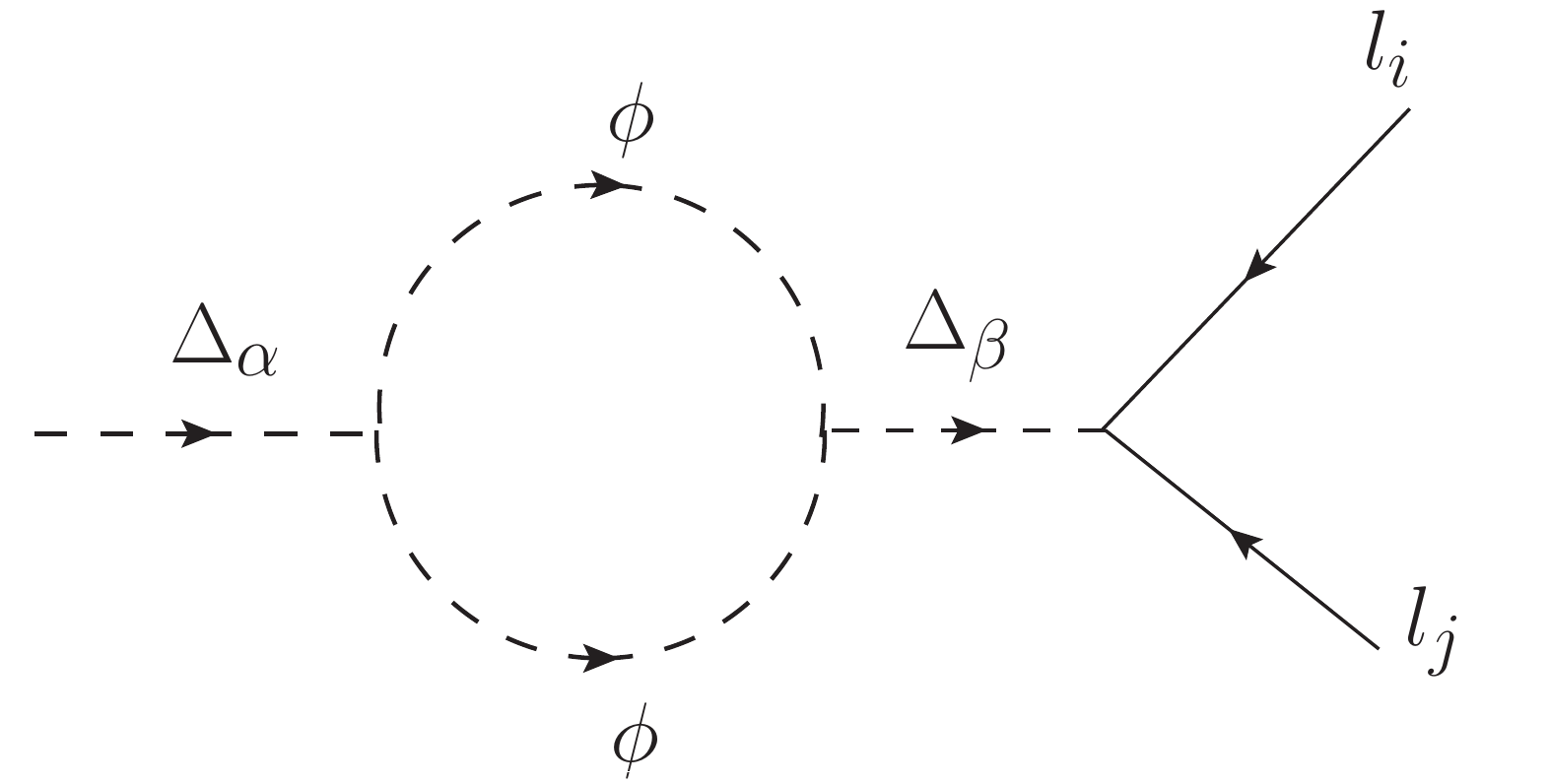}
 \includegraphics[width=5cm,height=3cm]{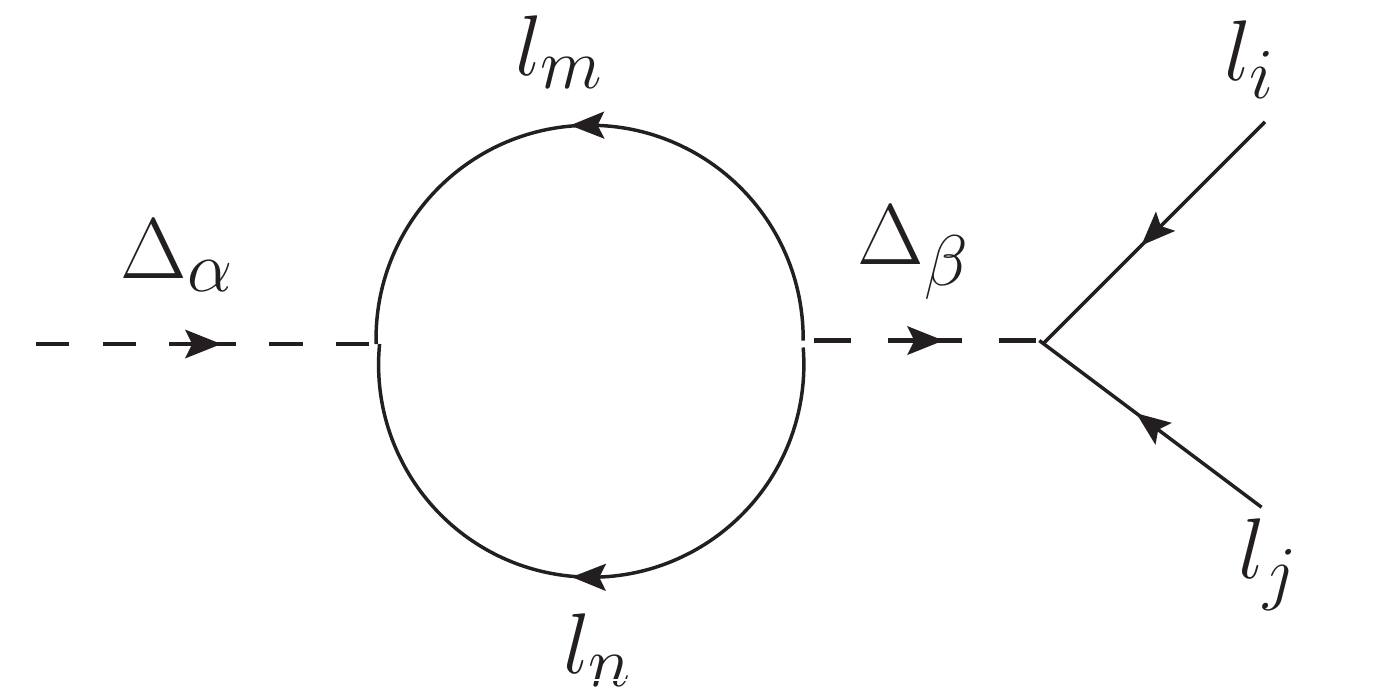}
 \caption{Tree level and one loop Feynman diagrams for a triplet decaying to bi-leptons in case of flavoured leptogenesis.}
 \label{feyn-flav}
 \end{center}
\end{figure}

 In the interaction 
Lagrangian in eq.(\ref{Yukhiggs}), lepton number violation (LNV)
occurs due to the coexistence of the Higgs triplet-bilepton Yukawa matrix $f$ along with the trilinear couplings 
$\mu_{\Delta_\alpha}(\alpha=1,2)$. In the  absence of heavy RHNs , the entire CP asymmetry is  due to the 
decay of the two heavy scalar triplets. At tree level the scalar triplets can decay to bileptons or to  SM Higgs with 
the corresponding branching ratios  
\begin{eqnarray}
&& B^\alpha_l=\sum_{i=e,\mu,\tau} B^\alpha_{l_{i}} = \sum_{i,j=e,\mu,\tau} B^\alpha_{l_{ij}} =\sum_{i,j=e,\mu,\tau} 
\frac{M_{\Delta_\alpha}}{8\pi \Gamma^{tot}_{\Delta_\alpha}} |f^{(\alpha)}_{ij}|^2 ~{\rm and}~\\
&& B^\alpha_\phi =\frac{|\mu_{\Delta_\alpha}|^2}{8\pi M_\Delta \Gamma^{tot}_{\Delta_\alpha}}, 
\end{eqnarray}
which  satisfy $B^\alpha_l+B^\alpha_\phi=1$. Here $\Gamma^{tot}_{\Delta_\alpha}$ is the total decay width of $\Delta_\alpha$
\begin{equation}
 \Gamma^{tot}_{\Delta_\alpha}=\frac{M_{\Delta_\alpha}}{8\pi}\Big( \sum_{i,j} |f^{(\alpha)}_{ij}|^2 +
 \frac{|\mu_{\Delta_\alpha}|^2}{M^2_{\Delta_\alpha}} \Big)~.
\end{equation}
The decay to 
bi-leptons also occurs due to loop  mediation caused by either the SM Higgs or the leptons as shown in 
Fig.\ref{feyn-flav}.
The CP-asymmetry which arises due to the interference of the tree level and one-loop contributions  consists of two pieces: (i) the scalar loop generated  asymmetry denoted by $\not L,\not F$  which violates both lepton number and flavour, (ii) the lepton loop generated  flavour only violating asymmetry denoted by $\not F$\cite{Sierra:2014tqa,pcns:2020}
\begin{equation}
\epsilon^{l_i}_{\Delta_\alpha}= \epsilon^{l_i(\not L,\not F)}_{\Delta_\alpha}+ \epsilon^{l_i(\not F)}_{\Delta_\alpha}~.
\label{flav_ep}
\end{equation}
The indices $n,i$ in the above expressions of CP asymmetries stand for the lepton flavour indices $(e,\mu,\tau)$. It is thus realised that when 
 summed over flavour indices, the second piece of the CP asymmetry arising  solely out of  flavour violation vanishes identically, i.e
\begin{equation}
 \sum_{i=e,\mu,\tau} \epsilon^{l_i(\not F)}_{\Delta} =0~.
\end{equation}
This CP asymmetry parameter survives only in the case of flavoured leptogenesis. Since this asymmetry does not involve any lepton number
violation, it is also called purely flavoured asymmetry and the corresponding leptogenesis scenario which is dominated by the 
flavour violating CP asymmetry $(\epsilon^{l_i(\not F)}_{\Delta} \gg \epsilon^{l_i(\not L,\not F)}_{\Delta})$ is referred to as 
purely flavoured (PFL) leptogenesis. The condition for PFL leptogenesis is \cite{Sierra:2014tqa}
\begin{equation}
\mu^\ast_{\Delta_2} \mu_{\Delta_1} \ll M^2_{\Delta_2} Tr(f^{(2)} {f^{(1)}}^\dagger)~.
\end{equation}
 Consistent with dominance of $m_{\nu}^{(2)}$ over  $m_{\nu}^{(1)}$ and the condition that $M_{\Delta_2} < M_{\Delta_1}\simeq \mu_{\Delta_1}$,  PFL condition has not been guaranteed in our flavoured leptogenesis regime \cite{pcns:2020}.

\par\noindent{\underline {\bf Leptogenesis guided  $\Delta_2-$ seesaw dominance:}}\\
Utilisation is made of the cosmological principle  that the decay of  the lower mass triplet $\Delta_2$ with $M_{\Delta_2}\ll M_{\Delta_1}$  generates the dominant asymmetry surviving the Hubble expansion rate of the universe as it erases out the asymmetry caused by the decay of the heavier triplet $\Delta_1$. This  leads to  the CP-asymmetry dominance by the decay of $\Delta_2$. More profoundly,  the inequality $M_{\Delta_2}\ll M_{\Delta_1}$ also predicts type-II seesaw dominance of neutrino mass matrix by $m_{\nu}^{(2)}$ mediated by $\Delta_2$. From the behaviour of two respective neutrino mass matrices  and
induced VEVs, 
$m_{\nu}^{(1)} \propto  V_{\Delta_1},\,
m_{\nu}^{(2)} \propto V_{\Delta_2},\,
V_{\Delta_1} \propto \frac{1}{M_{\Delta_1}^2},\,
V_{\Delta_2} \propto \frac{1}{M_{\Delta_2}^2}\,$,
 the leptogenesis guided principle  requiring $M_{\Delta_2}\ll M_{\Delta_1}$ predicts 
\begin{eqnarray}
m_{\nu}^{(1)}&\ll&m_{\nu}^{(2)},\nonumber\\
m_{\nu}&=&m_{\nu}^{(1)}+m_{\nu}^{(2)} \simeq m_{\nu}^{(2)} = m_{\nu}^{\rm DATA}.\label{eq:t2dom}
\end{eqnarray}        
This determines all the elements of the mass matrix $m_{\nu}^{(2)}$ from best fits to neutrino data given in Table \ref{tab:mnutab}. 

  For the sake of simplicity ignoring  the scalar triplet index and denoting
$\Delta_2\equiv \Delta$ result in
 $\epsilon_\Delta \equiv \epsilon_{\Delta_2},B_l \equiv B^2_l, B_\phi \equiv B^2_\phi$. Then the  two pieces of CP asymmetry
in eq.(\ref{flav_ep}) arising due to  
$\Delta_2$ decay are 
\begin{eqnarray}
&& \epsilon^{l_i(\not L,\not F)}_{\Delta} = \frac{1}{2 \pi} \frac{Im \left\{ \sum\limits_{n}^{} \left( f^{(2)} \right)^\ast_{ni} 
 \left({f^{(1)}}\right)_{ni} \mu^\ast_{\Delta_2} \mu_{\Delta_1} \right \}}{ M^2_{\Delta_2} Tr (  f^{(2)}  {f^(2)}^\dagger ) +|\mu_{\Delta_2}|^2 }g (x_{12}), \label{flav_ep1}\\
&& \epsilon^{l_i(\not F)}_{\Delta} = \frac{1}{2 \pi} \frac{Im \left\{ ( {f^{(2)}}^\dagger {f^{(1)}} )_{ii} Tr(f^{(2)} {f^{(1)}}^\dagger)  \right \}}{ M^2_{\Delta_2} Tr (  f^{(2)}  {f^(2)}^\dagger ) +|\mu_{\Delta_2}|^2 }g (x_{12}),\label{flav_ep2}
\end{eqnarray}
where  \\
\bea
 g(x_{\alpha\beta})&=&\frac{x_{\alpha\beta}(1-x_{\alpha\beta})}{(1-x_{\alpha\beta})^2 + x_{\alpha\beta}y }, \nonumber\\ 
x_{\alpha\beta}&=&\frac{M^2_{\Delta_\alpha}}{M^2_{\Delta_\beta}}, \nonumber\\
 y&=&\left( \frac{\Gamma^{tot}_{\Delta_\beta}}{M_{\Delta_\beta}} \right)^2.\label{eqg}
\eea
The indices $n,i$ in the above expressions of CP asymmetries stand for the lepton flavour indices $(e,\mu,\tau)$. It has been shown that purely flavoured leptogenesis \cite{Sierra:2014tqa} is not possible in the two-triplet seesaw model \cite{pcns:2020}.
The recent neutrino oscillation data are fitted using the leptogenesis guided $\Delta_2-$ seesaw dominance   with $M_{\Delta_2}\ll M_{\Delta_1}$ 
\beq
m_{\nu}\simeq m^{(2)}_{\nu}=m_{\nu}^{(DATA)}.  \label{eq:apmnu}
\eeq
Noting that the mass matrices $m_\nu^{(1)}$ and $m_\nu^{(2)}$  can always be represented as
\bea
(m_\nu^{(1)})_{ij}&=&|{m^{(1)}_\nu}_{ij} | e^{i\psi^{(1)}_{ij}} , \nonumber\\
 (m_\nu^{(2)})_{ij}&=&|{m^{(2)}_\nu}_{ij} | e^{i\psi^{(2)}_{ij}},\label{eq:mnu1mnu2}
\eea
gives
\bea
\frac{(m_\nu^{(1)})_{ij}}{(m_\nu^{(2)})_{ij}}&=& \frac{|{m^{(1)}_\nu}_{ij} |}{|{m^{(2)}_\nu}_{ij} |}   e^{i\phi_{ij} }~, \nonumber\\
&=& G_{ij} e^{i\phi_{ij}}. \label{eq:mnu12}
\eea
where
\bea
\phi_{ij}&=&\psi_{ij}^{(1)}-\psi_{ij}^{(2)}, \nonumber\\
G_{ij}&=& \frac{|(m^{(1)}_{\nu})_{ij} |}{|(m^{(2)}_{\nu})_{ij} |}.  \label{eq:Gphi} 
\eea
Then each element of $m_\nu^{(1)}$ is connected to the corresponding element of $m_\nu^{(2)}$ through a  multiplicative factor
which is a complex number
\begin{equation}
(m_\nu^{(1)})_{ij} =G_{ij} e^{i\phi_{ij}} (m_\nu^{(2)})_{ij} \label{mnu1_2}
\end{equation}
 
In the unflavoured regime the  lighter triplet mass has the lowest limit 
 $M_{\Delta_2} \gtrsim 4 \times 10^{11}$ GeV \cite{pcns:2020}.
Partially flavoured- or $\tau-$ flavoured leptogenesis (equivalent to $e+\mu$ flavoured-leptogenesis) has been found to be successful for $M_{\Delta_2} \simeq (10^{10}- 10^{11})$ GeV \cite{pcns:2020}.
In the unflavoured regime, the purely flavoured CP asymmetry part vanishes leading to non-vanishing lepton number plus flavour  violating
parts which have been represented in terms of the experimental value of $m_\nu(=m^{\rm DATA}_\nu)$,
\begin{eqnarray}
\epsilon^l_\Delta &=& \sum_i  \epsilon^{l_i(\not L,\not F)}_{\Delta} \nonumber\\
                &=& \frac{ M^2_{\Delta_1} M^2_{\Delta_2} }{2\pi v^4} 
                    \frac{\sum\limits_{ij} G_{ij} |(m_\nu)_{ij}|^2 \sin (\psi^{(1)}_{ij}-\psi^{(2)}_{ij}) }
                    {M^2_{\Delta_2} Tr (  f^{(2)}  {f^{(2)}}^\dagger ) +|\mu_{\Delta_2}|^2 }g (x_{12})  \label{unflav-asy1} \\                    
                 &\simeq& \frac{ M^2_{\Delta_1} M^2_{\Delta_2} }{16 \pi^2 v^4}    
                     \frac{\sum\limits_{ij} G_{ij} |(m_\nu)_{ij}|^2 \sin (\psi^{(1)}_{ij}-\psi^{(2)}_{ij}) }
                     { (M^2_{\Delta_1} -M^2_{\Delta_2}) } \left( \frac{M_{\Delta_2}}{\Gamma^{tot}_{\Delta_2}} \right) ~.\label{unflav-asy2} 
\end{eqnarray}
Then for  $M_{\Delta_1}\gg M_{\Delta_2}$, the CP-asymmetry is 
\bea
\epsilon^l_\Delta&=&\frac{ M^2_{\Delta_2} }{16 \pi^2 v^4}    
                     \sum\limits_{ij} G_{ij} |(m_\nu)_{ij}|^2 \sin (\psi^{(1)}_{ij}-\psi^{(2)}_{ij})  \left( \frac{M_{\Delta_2}}{\Gamma^{tot}_{\Delta_2}} \right) \nonumber\\ 
&\equiv& 
\frac{ M^2_{\Delta_2} }{16 \pi^2 v^4}    
                     \sum\limits_{ij} G_{ij}|(m_\nu)_{ij}|^2 \sin (\phi_{ij})  \left( \frac{M_{\Delta_2}}{\Gamma^{tot}_{\Delta_2}} \right) 
~.\label{unflav-asy3} 
\eea
 
Out of $9$ elements of the ratio $G_{ij}$, or the phase differences $\phi_{ij}$, only $6$ are independent which have been used as input to CP-asymmetry through the functions
\bea
F&=&(G_{11},G_{22},G_{33},G_{12},G_{13},G_{23}), \nonumber\\
\Phi&=&(\phi_{11},\phi_{22},\phi_{33},\phi_{12},\phi_{13},\phi_{23}). \label{eq:FG}
\eea
For the sake of simplicity, a common magnitude ratio has 
been fixed for different varieties of leptogenesis  solutions\cite{pcns:2020}
\beq
F=(0.1, 0.1, 0.1, 0.1, 0.1, 0.1), \label{eq:commag}
\eeq
which ensures $\Delta_2$ seesaw dominance.\\
 For phase differences between the corresponding elements of $m^{(1)}_\nu$ and $m^{(2)}_\nu$ we have used two different possibilities for  every numerical computation:\\
{\bf (i) Fixed phase difference:}\\ 
\begin{equation}
\Phi= (-\pi/2, -\pi/2, -\pi/2, -\pi/2, -\pi/2, -\pi/2),  \label{eq:fph}
\end{equation}
{\bf (ii) Randomised phase difference:}\\
\beq
\frac{\Phi}{\pi}= (-0.3418, -0.0807, 0.7850, 0.9961, -0.4427, 0.7244)~~  \label{eq:rph}
\eeq
where the assumption in eq.(\ref{eq:fph}) has been made to yield maximal CP-asymmetry as input to flavoured Boltzmann equations, but the more generalised phase differences of eq.(\ref{eq:rph})
has been suggested through random number generation \cite{pcns:2020}. 
Using the CP-asymmetry thus predicted by two-triplet seesaw model from neutrino data as input, 
the respective sets of coupled Boltzmann equations have been solved in different cases to predict the desired baryon asymmetry of the Universe \cite{pcns:2020}
defined through $Y_{B}$ or $\eta_B$ given below. 
\begin{equation}
 Y_B=\frac{n_B-n_{\overline{B}}}{s}.
\end{equation}
Here $n_B,n_{\overline{B}}$ are number densities of baryons and anti-baryons, respectively, and $s$ is the entropy density.
\begin{equation}
\eta_B=\frac{n_B-n_{\overline{B}}}{n_\gamma},
\end{equation}
 were $n_\gamma=$  photon density.
 Planck satellite experimental values  are\cite{Planck15}
\bea
&&Y_B=8.66\pm 0.11 \times 10^{-11},\label{YB}\\
&&\eta_B=6.10\pm 0.08 \times 10^{-10}.\label{etaB}
\eea

Our estimations of BAU \cite{pcns:2020} have been found to match with the Planck satellite experimental values given in eq.(\ref{YB}) and eq.(\ref{etaB}).
The masses of triplets, trilinear couplings, types of neutrino mass hierarchy and phase difference $\Phi/\pi$ which have yielded successful predictions of BAU  used  and the corresponding leptogenesis type in each case have been summarised  in Table \ref{tab:masspara}. 
\begin{table}[!h]
\caption{Scalar triplet masses and lepton number violating couplings of the two-triplet model leading to baryon asymmetry prediction through leptogenesis in concordance with current neutrino data and cosmological bounds for NO and IO type mass hierarchies \cite{pcns:2020}.}
\label{tab:masspara}
\begin{center}
\begin{tabular}{|c|c|c|c|c|c|c|}
\hline
$M_{\Delta_1}$ &$\mu_{\Delta_1}$&$M_{\Delta_2}$&$\mu_{\Delta_2}$&Hierarchy& Leptogenesis&Phase\\
${(\rm GeV)}$&${(\rm GeV)}$&${(\rm GeV)}$&${(\rm GeV)}$&type&type&difference\\
\hline
\hline
$3\times 10^{13}$&$10^{13}$&$10^{12}$&$2.4 \times 10^{10}$&NO&Unflavoured&Fixed\\
\hline
$3\times 10^{13}$&$10^{13}$&$ 10^{12}$&$2.0\times 10^{10}$&IO&Unflavoured&Fixed\\
\hline 
$3\times 10^{13}$&$10^{13}$&$5\times 10^{12}$&$2.6\times 10^{11}$&NO&Unflavoured&Random\\
\hline
$3\times 10^{13}$&$10^{13}$&$5\times 10^{12}$&$2.1\times 10^{11}$&IO&Unflavoured&Random\\
\hline
$3\times 10^{12}$&$10^{12}$&$10^{11}$&$3.6\times 10^{9}$&NO&$\tau$-flavoured&Random\\
\hline
$3\times 10^{12}$&$10^{12}$&$10^{11}$&$3.2\times 10^{9}$&IO&$\tau-$flavoured&Random\\
\hline
\hline 
\end{tabular}
\end{center}
\end{table}

\par In addition to the masses of the two scalar triplets given in Table \ref{tab:masspara},
we  have checked that partially flavoured or two-flavoured ($\equiv \tau-$ flavoured ) leptogenesis is also possible with desired BAU prediction 
for mass parameters $M_{\Delta_1}\simeq {\cal O}(10^{11})$ GeV, $M_{\Delta_2}\simeq {\cal O}(10^{10})$ GeV, $\mu_{\Delta_1}\simeq {\cal O}(10^{10})$ GeV, and $\mu_{\Delta_2}\simeq {\cal O}(10^{9})$ GeV with NO type neutrino mass hierarchy and randomised
values of the phase differences. In particular this lighter $\Delta_2$  mass $M_{\Delta_2} \sim {\cal O}(10^{10})$ GeV (not shown in Table \ref{tab:masspara}) is approximately estimated lowest bound on the scale for viable $\tau-$ flavoured leptogenesis. 
\section{Unification through SU(5) }\label{sec:su5}
\subsection{Lighter fermions and scalars}\label{sec:lfs}

Embedding  the two-Higgs triplets  in  SO(10) (or $E_6$)  GUT would result in  additional contributions due to RHN mediation for both neutrino masses and BAU via leptogenesis which could destroy  the special property 
 of the
 TTM that both neutrino masses and  leptogenesis need no RHNs \cite{Ma-Us:1998,pcns:2020}. 
 On the other hand  all the standard fermions of one  generation  are contained in two fundamental SU(5) representations, ${\overline 5}_F \oplus {10}_F$, whose decompositions under SM gauge theory $G_{213}$ are
\bea
{\overline 5}_F&\supset& l_L(2,-1/2,1)+d_R(1,-1/3,3), \nonumber\\
{10}_F&\supset& Q_L(2, 1/6, 3)+u_R(1, 2/3, 3)+e_R(1, -1, 1). \label{eq:su5ff}
\eea   
 Thus like the SM, the 
minimal SU(5) theory does not have RHNs among its nontrivial fermion representations. This qualifies SU(5) as a suitable framework to unify the two triplet model.
Another explicit advantage over  GUTs of higher rank  is due to much smaller size of the required SU(5) Higgs representations ${15}_{H1}\supset \Delta_1$ and ${15}_{H2}\supset \Delta_2$. Noting that the same two triplets are contained in the two scalar representations ${126}_{H1}\oplus {126}_{H2}$ of SO(10)( or ${351}^{\prime}_{H1}\oplus {351}^{\prime}_{H2} \subset E_6$), it is evident that the two scalar representations 
needed in SU(5) for the TTM embedding are much smaller compared to the corresponding representations of SO(10) (or $E_6$).\\
Type-II seesaw and proton decay in SU(5) using ${15}_H$ have been addressed \cite{DP:2005} which need lepto-quark scalar masses ${\cal O} (1)$ TeV. As discussed below, such scalar
leptoquarks in ${15}_{H1}$ of the present TTM embedding  having masses $\gg 10^{10}$ GeV, do not make any substantial  contribution to proton decay \cite{cps:2019,cps:AHEP:2018}. 
 As  SU(5)  does not permit any  intermediate gauge symmetry \cite{cmp:1984}, gauge coupling unifications have been suggested  by populating the grand desert through non-standard fermions or scalars, or both types of fields 
\cite{Perez:2019-1,Perez:2018-2,RNM-mkp:2011,Bajc-Nem-gs:2007,Ma-Suematsu:2008,Frig-Ham:2010,psb:2010,mkp:2011,Perez-1,Bajc-gs:2007,Kynshi-mkp:1993}. 
The present gauge coupling unification method must be  through inclusion of possible additional fields in such a way that the heavy masses of two triplets
$\Delta_1$ and $\Delta_2$ are separated by at least one order  \cite{pcns:2020}.        
For later use in this section we decompose the scalar  (H) and  fermion (F) representations of the SU(5) under $G_{213}$ gauge group \cite{Slansky:1979}
\par\noindent{\bf $SU(5) \supset SU(2)_L\times U(1)_Y\times SU(3)_C (\equiv G_{213})$:}
\bea
{5}_H&\supset& \phi (2,1/2,1)+(1,-1/3,3),\nonumber\\
{15}_{H1}&\supset&\Delta_1(3, -1, 1)+ (2,1/6,3)+(1,2/3,6),\nonumber\\ 
{15}_{H2}&\supset&\Delta_2(3, -1, 1)+ (2,1/6,3)+(1,2/3,6).\nonumber\\ 
{24}_H&\supset& S_{24}(1,0, 1)+(1,3,0)+(1,0,8)+ (2,-5/6, 3)+(2, 5/6, {\overline 3}),\nonumber\\ 
{24}_F&\supset&\Sigma(3, 0,1)+C_F(1,0,8)_F+S_F(1,0,1)_F\nonumber\\
&+&{(\rm LQ)}_F(2,-5/6,3)_F+{\overline {\rm LQ}}_F(2,5/6,{\bar 3})_F.\label{eq:dec}
\eea
We have embedded the two Higgs triplets $\Delta_1$ and $\Delta_2$ in the SU(5) representations ${15}_{H1}$ and ${15}_{H2}$, respectively. Here $\phi (2, 1/2,1) \subset {5}_H$ is the standard Higgs doublet. The scalar singlet $S_{24}\subset {24}_H$ is usually exploited to break SU(5) $\to$ SM. Extended survival hypothesis applies under minimal fine tuning \cite{del Aguilla:1981,RNM-gs:1983}  to maintain  gauge hierarchy resulting in only the SM Higgs doublet $\phi$ to be at the electroweak scale. All other scalars not needed
  for electroweak symmetry breaking acquire masses at the GUT scale $M_U$ under minimal fine tuning hypothesis \cite{del Aguilla:1981,RNM-gs:1983}.
But since two lighter scalar triplets are needed  for neutrino mass and BAU ansatz \cite{pcns:2020}, additional fine tuning is needed for such realizations. However, the present unification model is  found to operate if 
all the component masses of the complete  multiplet ${15}_{H1}$ can be identified with the heavier triplet mass scale $M_{15_{H1}}=M_{\Delta_1}$, an interesting criteria which has been exploited for unification in SUSY \cite{Goh-RNM-Nasri:2004},
split-SUSY \cite{RNM-mkp:2011} and non-SUSY \cite{mkp-rs:2020,cps:2019} cases. For implementation of this criteria it is necessary to realize unification of SM gauge couplings with all other necessary particle masses lighter than $M_{\Delta_1}=M_{15_{H1}}$ \cite{Goh-RNM-Nasri:2004}. 
In a straightforward manner the mass of $\Delta_2\subset {15}_{H_2}$ can be made lighter than the GUT scale by using the SU(5) invariant potential 
\bea
V_{15H2}&=&M_{15_{H2}}^2{15}_{H2}^{\dagger}{15}_{H2}+\lambda_{15H2}({15}_{H2}^\dagger{15}_{H2})^2+m_{(24,15H2)}{15}_{H2}^{\dagger}{24}_H{15}_{H2}    \nonumber\\
&+&\gamma_2Tr({24}_{H}^2){15}_{H2}^{\dagger}{15}_{H2}+\delta_2{15}_{H2}^\dagger{24}_H^2{15}_{H2}+m_{(15H2,5_H)}{15}_{H2}5_H5_H.....\label{eq:V15H2}  
\eea
When the symmetry breaking SU(5) $\to$ SM occurs through  VEV
\beq
\langle S_{24H} \rangle =\frac{V_U}{\sqrt 30}diag(2,2,2,-3,-3), \label{eq:vevu}
\eeq 
we get for the mass squared term for $\Delta_2$ 
\beq
M_{\Delta_2}^2=M_{15_{H2}}^2-\frac{3m_{(24,15H2)}V_U}{\sqrt 30}+\frac{3}{10}\lambda_{(24,15H2)}V_U^2+\gamma_2V_U^2+\frac{3}{10}\delta_2 V_U^2, \label{eq:MDEL2}
\eeq
where $M_{15_{H2}}, m_{(24,15H2)}$  and $V_U$ are   $\sim {\cal  O}(M_U)$. Then it is
clear that there is ample scope to make $M_{\Delta_2}^2\ll M_U^2$ by fine tuning of parameters in  eq.(\ref{eq:MDEL2}).\\
In Sec. \ref{sec:numass} the LNV couplings occurring in eq.(\ref{Yukhiggs}) originate from the two trilinear SU(5) invariant couplings, ~~$m_{(15Hi,5_H)}{15}_{Hi}5_H5_H \supset \mu_{\Delta_i}\Delta_i\phi\phi (i=1,2)$ whose numerical values have been shown in Table \ref{tab:masspara}, Table \ref{tab:RGEsol}, and Table \ref{tab:RGEsol2}.\\
Besides the HIggs scalar triplets needed for neutrino mass and leptogenesis, as we show below, completion of gauge coupling unification  also needs the fermion triplet $\Sigma(3,0,1) \subset {24}_F$ and the colour octet fermion $C_F(1,0,8) \subset {24}_F$ to be substantially lighter than the GUT scale with masses $M_{\Sigma}\sim {\cal O}(500-3000)$ GeV and $M_{C_F}\sim {\cal O}(10^8-10^9)$ GeV, respectively.  For making non-standard fermions $\Sigma$ and $C_F$ lighter, we utilise the most convenient  Yukawa Lagrangian  including both normalizable as well as non-renormalizable (NR) terms \cite{Bajc-gs:2007}
\bea
{\cal L}_{NR}&=&M_{F}Tr({24}_F^2) + Y_{24}Tr({24}_F^2{24}_H \nonumber\\
 &+&\frac{1}{M_{NR}}\left (k_1Tr({24}_F^2)Tr({24}_H^2)+ k_2[Tr({24}_F{24}_H)]^2\right) \nonumber\\
&+&\frac{1}{M_{NR}}\left (k_3Tr({24}_F^2{24}_H^2)+k_4Tr({24}_F{24}_H{24}_F{24}_H)]
\right ).  \label{eq:GYUK}
\eea 
Whereas all the four kinds of non-standard fermion masses such as the singlet,
the leptoquarks, the weak triplet, and the colour octet fermion specified in eq.(\ref{eq:dec}) were needed to be lighter than GUT scale
in \cite{Bajc-gs:2007}, we need only the fermion triplet $\Sigma$ and the colour octet fermion $C_F$ to be lighter. Therefore, compared to \cite{Bajc-gs:2007} we need to fine tune relatively less number of parameters.    
For the mass scale in the non-renormalisable term, we use $M_{NR} \simeq
$ string scale $= M_{String}\sim 10^{17}$ GeV (or reduced Planck scale $\simeq 10^{18}$ GeV). When  the Yukawa couplings are switched off, all component masses in ${24}_F$ are near the GUT scale with
 degenerate masses $=M_F$. When SU(5) is broken by the VEV given in eq.(\ref{eq:vevu})
with  $V_U\sim M_{GUT}$, masses of  $\Sigma(3,0,1)$ and  $C_F(1,0,8)$ are
\bea
M_{\Sigma}&=&M_{F}-\frac{3Y_{24}V_U}{\sqrt 30}+\frac{V_U^2}{M_{NR}}\left(k_1+\frac{3}{10}(k_3+k_4)\right),\nonumber\\
M_{C_F}&=&M_F+\frac{2Y_{24}V_U}{\sqrt 30}+\frac{V_U^2}{M_{NR}}\left(k_1+\frac{2}{15}(k_3+k_4)\right) \label{eq:FMASS}
\eea
Other components in ${24}_F$  can be also expressed in  a similar fashion as discussed in the Appendix. As we discuss below, our RG constraints on gauge coupling  unification need  $\Sigma$ whose masses can be in the range $M_{\Sigma} \sim {\cal O} (500-3000)$ GeV, and correspondingly $M_{C_F}\sim {\cal O} (10^8-10^9)$ GeV compared to GUT scale values of $M_F\sim V_U \sim 10^{16}$ GeV. Then  with appropriate fine tuning of parameters, the two relations in eq.(\ref{eq:FMASS}) can give these desired masses when 
\bea
k_3&+&k_4 \sim 0, \nonumber\\
M_{F}&-&\frac{3Y_{24}V_U}{\sqrt 30}+\frac{V_U^2}{M_{NR}}k_1\sim 0,\\
M_{C_F}&=&\frac{5Y_{24}V_U}{\sqrt 30} \label{eq:FMsol} 
\eea
for $Y_{24}\sim 10^{-7}-10^{-5}$. When these relations are used, all  component masses become superheavy except $\Sigma$ and $C_F$ as discussed in more detail in the Appendix in Sec.\ref{sec:A1}.
\section{Gauge coupling unification}\label{sec:uni}
\subsection{\textbf{ Unification with lighter fermions and scalars}}
As discussed above, for successful unification with $\Delta_1,\Delta_2$ at two different heavy mass scales, the weak fermion triplet $\Sigma(3,0,1)$ and the colour octet fermion  $C_F (1,0,8)$ are necessary. The mass hierarchy of  different fields contributing to  gauge coupling unification is arranged in the following order
\beq
M_U \ge M_{\Delta_1}( =M_{15_{H1}})\gg M_{\Delta_2} \gg M_{C_F}\gg M_{\Sigma} \gg M_W, \nonumber\\
\eeq   
We use the standard renormalization group equations (RGEs) for the evolution
 of gauge
 couplings \cite{GQW:1974, Jones:1982,cmgmp:1985,Lang-Pol:1993,mkp-Cajee:2005}
 and the  analytic solutions
are 
\begin{eqnarray}
\frac{1}{\alpha_i(M_Z)}&=&\frac{1}{\alpha_i(M_{U})}
+\frac{a^{(1)}_i}{2\pi}{\rm ln}\left(\frac{M_{\Sigma}}{M_Z}\right)
+\frac{a^{(2)}_i}{2\pi}{\rm ln}\left(\frac{M_{C_F}}{M_{\Sigma}}\right)
+\frac{a^{(3)}_i}{2\pi}{\rm ln}\left(\frac{M_{\Delta_2}}{M_{C_F}}\right) \nonumber\\ 
&+&\frac{a^{(4)}_i}{2\pi}{\rm ln}\left(\frac{M_{\Delta_1}}{M_{\Delta_2}}\right) 
+\frac{a^{(5)}_i}{2\pi}{\rm ln}\left(\frac{M_{U}}{M_{\Delta_1}}\right)
\nonumber\\  
&+&\Theta^{(1)}_i+\Theta^{(2)}_i+\Theta^{(3)}_i +\Theta^{(4)}_i+\Theta^{(5)}_i, (i=Y,2L,3C), \label{tlrge}
\end{eqnarray}
In fact unification is at first realized \cite{psna:2017} without $\Delta_1\subset {15}_{H1}$ which is then superimposed along with all other components of ${15}_{H1}$  upon the RG evolution pattern with $M_{\Delta_1}=M_{15_{H1}}\ge 10 M_{\Delta_2}$.
Precision gauge coupling unification  is found to be left unaltered
by such superimposition except for changing the value of $\alpha_G^{-1}$ by $\le 5\%$ \cite{cps:2019,Goh-RNM-Nasri:2004,RNM-mkp:2011}.
 The one-loop and two-loop 
 coefficients  in their respective ranges of mass
 scales are presented in Table \ref{tab:beta}. The terms $\Theta^{(k)}_i, (k=1,2,...,5)$\,\, are the two-loop contributions in the respective
  ranges of  mass scales in this unification model.
Defining the four two-loop to one-loop beta function coefficient ratios
\beq
B^{(k)}_{ij}=a^{(k)}_{ij}/a^{(k)}_j (k=1,2,3,4), \label{tocr}
\eeq
the $\Theta^{(k)}_i$ functions are
\begin{eqnarray}
&&\Theta^{(1)}_i=\frac{1}{4\pi}\sum_j B^{(1)}_{ij}{\rm ln}\frac{\alpha_j(M_{\Sigma})}{\alpha_j(M_Z)},
~~\Theta^{(2)}_i=\frac{1}{4\pi}\sum_j B^{(2)}_{ij}{\rm ln}\frac{\alpha_j(M_{C_8})}{\alpha_j(M_{\Sigma})},\nonumber\\
&&\Theta^{(3)}_i=\frac{1}{4\pi}\sum_j B^{(3)}_{ij}{\rm ln}\frac{\alpha_j(M_{\Delta})}{\alpha_j(M_{C_8})},\nonumber\\
&&\Theta^{(4)}_i=\frac{1}{4\pi}\sum_j B^{(4)}_{ij}{\rm ln}\frac{\alpha_j(M_{\Delta_1})}{\alpha_j(M_{\Delta_2})}. \label{thdef}
\end{eqnarray}
As $a_2^{(5)}=0$, $\Theta_i^{(5)}$ is determined by direct integration in the range
$M_{\Delta_1}\to M_U$
\beq
\Theta^{(5)}_i=\frac{1}{8\pi^2}\sum_j a^{(5)}_{ij}{\int}_{M_{\Delta_1}}^{M_U}\alpha_j\frac{d\mu}{\mu}.                          \label{eq:th5int}
\eeq

The particle content in  different ranges of mass scales are summarised in Table \ref{tab:particles}.
 For the respective beta functions the one-loop  coefficients $a^{(k)}_i (k=1,2,..,5)$, and  
the  two-loop coefficients $a^{(k)}_{ij} (k=1,2,..,5)$ are given in Table \ref{tab:beta}. \\
\begin{table}[h!]
\centering
\begin{tabular}{|l|l||l|}
\hline 
Energy Scale & Particle content  \\
\hline
$M_Z-M_{\Sigma}$ & SMPs   \\ 
\hline
$M_{\Sigma}-M_{C_F}$ & SMPs$+\Sigma$ \\
\hline
$M_{C_F}-M_{\Delta_2}$ &  SMPs$+\Sigma +C_F$ \\
\hline
$M_{\Delta_2}- M_{\Delta_1}$ & SMPs$+\Sigma+C_F +\Delta_2$ \\
\hline
$M_{\Delta_1}-M_{U}$& SMPs$+\Sigma+C_F+\Delta_2+{15}_{H1}$\\
\hline
\hline
\end{tabular}
\caption{Particle content of the model in different ranges of mass scales
 where SMPs stand for standard model particles.}
\label{tab:particles}
\end{table}
\begin{table}[h!]
\centering
\begin{tabular}{|c|c|c|c|c|}
\hline
\hline
Mass range and k & $a^{(k)}_i$ & $a^{(k)}_{ij}$  \\
\hline
$M_Z-M_{\Sigma},(k=1)$&$\bpmat 41/10 \\ -19/6 \\ -7\epmat$ & 
$\bpmat
199/50 &27/10&44/5\\
9/10 &35/6 & 12\\
11/10& 9/2 &-26\\
\epmat$
\\ 
\hline
$M_{\Sigma}-M_{C_F}, (k=2)$ &$\bpmat 41/10 \\ -11/6 \\ -7\epmat$ & 
$\bpmat
199/50 &27/10&44/5\\
9/10 &163/6 & 12\\
11/10& 9/2 &-26\\
\epmat$
\\ 
\hline
$M_{C_F}-M_{\Delta_2}, (k=3)$ & $\bpmat 41/10 \\ -11/6 \\ -5\epmat$ & 
$\bpmat
199/50 & 27/10 & 44/5\\
9/10 & 163/6 & 12\\
11/10& 9/2 & 22\\
\epmat$
 \\ 
\hline
$M_{\Delta_2}-M_{\Delta_1}, (k=4)$ & $\bpmat 47/10 \\ -7/6\\ -5\epmat$ & 
$\bpmat
83/10 & 171/10 & 44/5\\
57/10 & 275/6 & 12\\
11/10& 9/2 & 22\\
\epmat$
\\ 
\hline
$M_{\Delta_1}-M_{U}, (k=5)$ & $\bpmat 88/15 \\ 0 \\ -23/6\epmat$ & 
$\bpmat
641/50 & 43/2 & 44/5\\
21/2 & 387/6 & 12\\
11/10& 9/2 & 22\\
\epmat$
\\
\hline
\hline
\end{tabular}
\caption{One-loop and two-loop beta function coefficients in the respective ranges of
mass scales.}
\label{tab:beta}
\end{table}

Under the present TTM unification program through SU(5), confining for the sake of simplicity to predominantly one loop RG contributions, we have tested the model capabilities for precision gauge coupling unification for each input value of triplet fermion mass encompassing the range $M_{\Sigma}\simeq {\cal O}(500-3000)$ GeV consistent with neutrino oscillation data, cosmological bounds on neutrino masses and heavy scalar triplet mass predictions necessary for unflavoured or $\tau-$ flavoured leptogenesis. For this purpose we note that in Table \ref{tab:masspara} there are three distinct sets of  mass-parameter
 combinations, $(M_{\Delta_1}, M_{\Delta_2}, \mu_{\Delta_1})$, each set leading to two
solutions corresponding to NO and IO type neutrino mass hierarchies \cite{pcns:2020} depending upon two different values of the LNV coupling parameter $\mu_{\Delta_2}$.
\par\noindent{\bf Set-I:}
\bea
M_{\Delta_1}&=&3\times 10^{12} {\rm GeV},\nonumber\\ 
\mu_{\Delta_1}&=&m_{(15H1,5_H)}= 10^{12} {\rm GeV}, \nonumber\\
M_{\Delta_2}&=& 10^{11} {\rm GeV}, \nonumber\\
\mu_{\Delta_2}&=& m_{(15H2,5_H)}=3.6\times 10^{9} {\rm GeV},\,\,(\rm NO), \nonumber\\
\mu_{\Delta_2}&=& m_{(15H2,5_H)}=3.2\times 10^{9} {\rm GeV}, \,\, (\rm IO). \label{eq:set1}
\eea
\par\noindent{\bf Set-II:}
\bea
M_{\Delta_1}&=&3\times 10^{13} {\rm GeV},\nonumber\\ 
\mu_{\Delta_1}&=&m_{(15H1,5_H)}= 10^{13} {\rm GeV}, \nonumber\\
M_{\Delta_2}&=&10^{12} {\rm GeV}, \nonumber\\
\mu_{\Delta_2}&=&m_{(15H2,5_H)}= 2.4\times 10^{10} {\rm GeV}, \,\,(\rm NO), \nonumber\\
\mu_{\Delta_2}&=&m_{(15H2,5_H)} 2.0\times 10^{10} {\rm GeV}, \,\,(\rm IO). \label{eq:set2}
\eea
 \par
\begin{figure}
\begin{center}
\includegraphics[width=10cm,height=8cm]{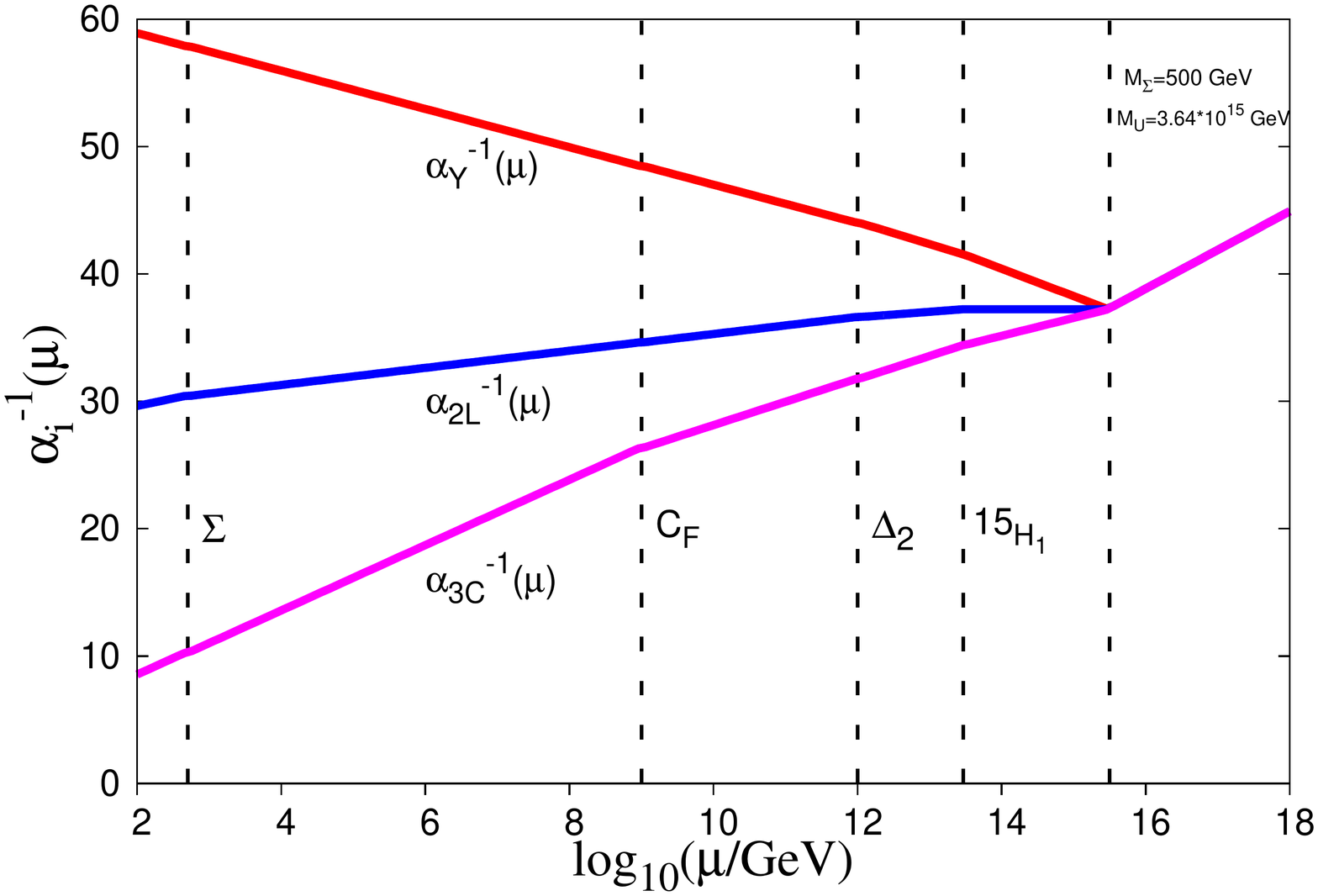}
\caption{ Unification of inverse fine structure constants of the SM gauge group
  $SU(3)_C\times SU(2)_L\times U(1)_Y$ at the GUT scale $M_U=10^{15.54}$ GeV in the presence of masses of lighter scalar triplet
    $M_{\Delta_2}=10^{12}$ GeV,  heavier scalar triplet $M_{\Delta_1}=M_{{15}_H}=10^{13.5}$ GeV, weak triplet fermion   $M_{\Sigma}=500$ GeV,  and colour octet fermion $M_{C_F}=10^9$ GeV.}
\label{fig:invf}
\end{center}
\end{figure}

\begin{figure}
\begin{center}
\includegraphics[width=10cm,height=8cm]{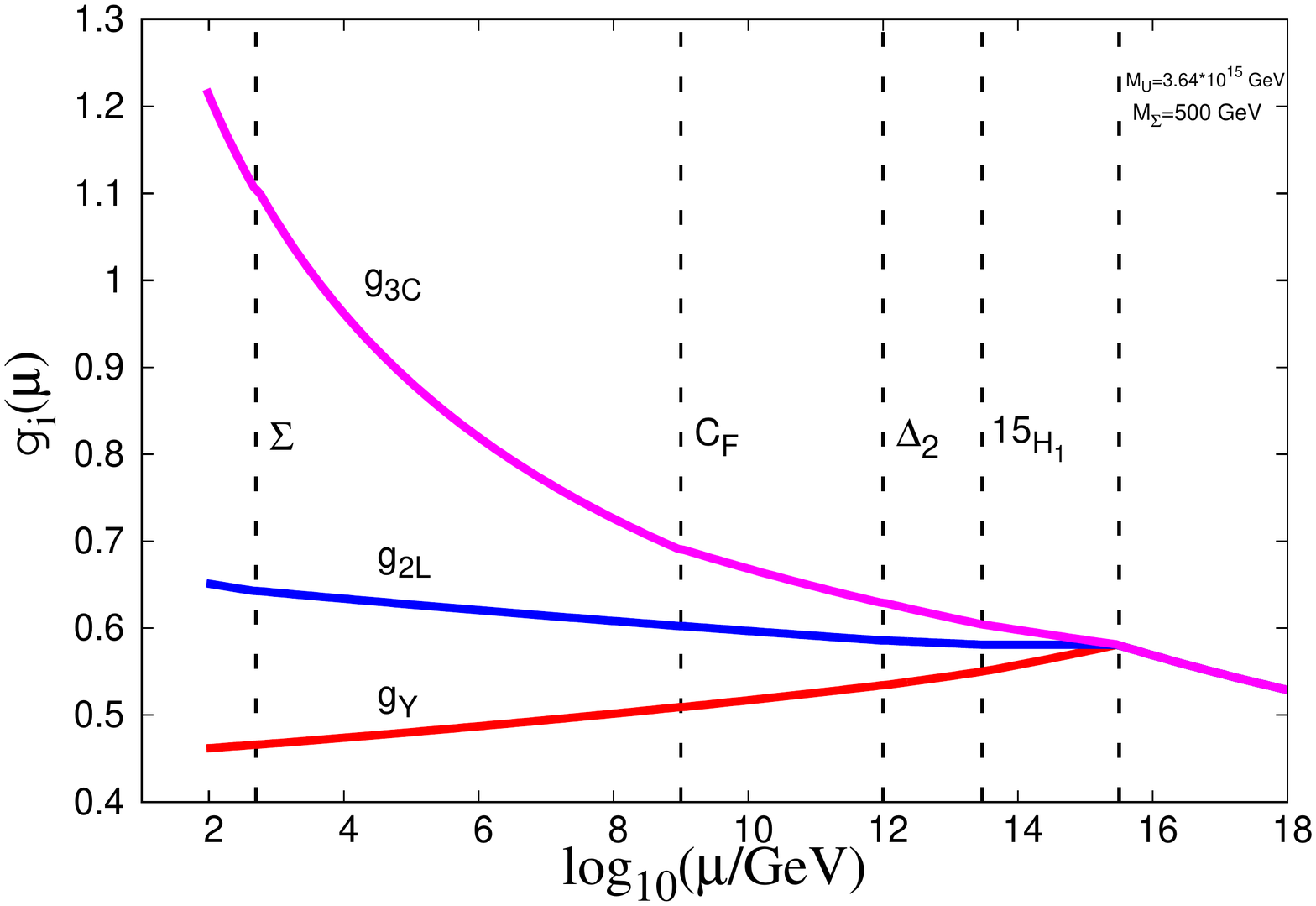}
\caption{Renormalization group evolutions of three gauge couplings
of the SM in the presence of two scalar triplets and other particles as in Fig. \ref{fig:invf} showing precision unification at $\mu=M_U=10^{15.54}$ GeV for $M_{\Sigma}=500$ GeV.}               
\label{fig:gi}
\end{center}
\end{figure}

\noindent{\bf Set-III:}
\bea
M_{\Delta_1}&=&3\times 10^{13} {\rm GeV},\nonumber\\ 
\mu_{\Delta_1}&=&m_{(15H1,5_H)}= 10^{13} {\rm GeV}, \nonumber\\
M_{\Delta_2}&=&5\times 10^{12} {\rm GeV}, \nonumber\\
\mu_{\Delta_2}&=&m_{(15H2,5_H)}= 2.6\times 10^{11} {\rm GeV},\,\,(\rm NO),  \nonumber\\
\mu_{\Delta_2}&=&m_{(15H2,5_H)}= 2.1\times 10^{11} {\rm GeV},\,\,(\rm IO).  \label{eq:set3}
\eea
 where we have identified in this work the respective LNV couplings
 $\mu_{\Delta_i}(i=1,2)$  as originating from fine tuning of respective SU(5) invariant trilinear couplings 
\beq
\mu_{\Delta_i}=m_{(15Hi,5_H)}\,\,(i=1,2)\subset m_{(15Hi,5_H)} \left(15Hi.{5}_H {5}_H\right )\,\, (i=1,2), \label{eq:tri}
\eeq
as also noted through eq.(\ref{eq:V15H2}) and discussions following 
eq.(\ref{eq:MDEL2}). As these trilinear coupling values  do not affect running of gauge couplings, it is sufficient to prove precision gauge coupling unification of all the six different solutions of Table \ref{tab:masspara} if we prove such unification only for three different sets of scalar mass triplet combinations $(M_{\Delta_1}, M_{\Delta_2})$ given in eq.(\ref{eq:set1}), eq.(\ref{eq:set2}), and eq.(\ref{eq:set3}). In our TTM unification program through SU(5) presented below in Table \ref{tab:RGEsol} and Table \ref{tab:RGEsol2} we have designated such solutions as Sol.-I, Sol.-II, and Sol.-III each of which  represents the respective pair of solutions of Table \ref{tab:masspara} and eq.(\ref{eq:set1}), eq. (\ref{eq:set2}), and eq.(\ref{eq:set3}). To be more explicit, each of the unification solutions (Sol.-I, Sol.-II, and Sol.-III) presented in Table \ref{tab:RGEsol} or Table \ref{tab:RGEsol2} has the two sets of values of trilinear couplings specified in eq.(\ref{eq:set1}), eq.(\ref{eq:set2}), and eq.(\ref{eq:set3}) corresponding to the respective  pair of solutions of Table 3 representing NO or IO type neutrino mass hierarchy. Thus every set of parametric solutions of Table \ref{tab:masspara} derived in \cite{pcns:2020} has one-to-one correspondence to our unification solutions realised in this work.    \\
It has been theoretically concluded  that the thermally produced triplet fermion  $\Sigma$ can act as the whole of the
DM of the Universe by accounting for the entire value of the observed   cosmological   relic density $(\Omega h^2)_{Obs}=0.1172-0.1224$ \cite{Planck15,wmap}$=
(\Omega h^2)_{\Sigma}$ if its perturbatively estimated resonance mass is $M_{\Sigma}\simeq 2.4$ TeV \cite{Ma-Suematsu:2008} which increases to $M_{\Sigma} \ge 2.74$ TeV when non-perturbative Sommerfeld enhancement  is included \cite{Cirelli:2006,Frig-Ham:2010}. This non-perturbative resonance value has been also noted to be heavier, $M_{\Sigma}\simeq 3.0-3.2$ TeV \cite{Hryzcuk:2014}, or even $M_{\Sigma}\simeq 4$ TeV \cite{Mohanty:2012}. All such heavier resonance masses of the fermionic triplet as dominant thermal
DM are certainly accommodated by one class (designated under class (i) solutions) of our RG solutions to the present TTM unification program through SU(5)\footnote{We have checked that similar Unification solutions are possible also with $M_{\Sigma}\simeq 4 $ TeV \cite{Mohanty:2012} leading to $M_U\simeq {\cal O} (10^{15})$ GeV.}.\\  
On the other hand, we have discussed in Sec. \ref{sec:sigind} and Sec. \ref{sec:fxidm} a number of indirect experimental search constraints which forbid $\Sigma$ to have such heavier resonance mass values.  If these indirect search constraints are respected, then
for lighter $M_{\Sigma}\ll 2.4$ TeV, permitted by the other class ($=$ class (ii)) of the present RG unification solutions, $\Sigma$ may act as a sub-dominant  thermal DM component that can not account for the entire value of the observed relic density. In that case, for such class (ii) solutions, as we discuss below, we introduce an additional  scalar singlet DM component  $\xi$. The combined scalar and fermionic component contribution is suggested to account for the observed relic density as discussed in Sec. \ref{sec:fxidm}. Our RG solutions for TTM unification through SU(5) is found to accommodate both these two classes of WIMP DM solutions: (i) dominant fermionic triplet DM solutions with $M_{\Sigma}\ge 2.4$ TeV where a resonance value of $M_{\Sigma}$ accounts for the observed relic density  (ii) sub-dominant lighter fermionic triplet DM with $M_{\Sigma} \simeq {\cal O}(500-2000)$ GeV permitted by indirect search constraints where the added presence of a scalar singlet DM ${\xi}$ compensates for the deficit in the observed value of relic density.\\
Renormalisation group (RG) evolution of three inverse fine structure constants for one of these 
lower $\Sigma$ masses under the class (ii) model solutions with $M_{\Sigma}=500$ GeV
 is shown in Fig.\ref{fig:invf}.
We have also examined the RG evolutions of the three individual gauge couplings
of the SM, $g_i (\mu)\,\,(i=Y, 2L, 3C)$ which have been shown in Fig. \ref{fig:gi}. 

 The colour octet fermion masses needed for unification are
safely above the cosmologically allowed limit \cite{Arkani:2004}.
 Our predictions of mass scales and the GUT scale gauge coupling are
also presented in Table \ref{tab:RGEsol} as an example of class (ii) solutions 
for lighter $\Sigma$ with $M_{\Sigma}=500$ GeV.   

\begin{table}[h!]
\centering
\begin{tabular}{|c|c|c|c|c|c|c|c|c|}
\hline
\hline
Sol.&$\alpha_G^{-1}$&$M_U$&$M_{C_F}$&$M_{\Delta_1}$&$m_{(15H1,5_H)}$&$M_{\Delta_2}$&$m_{(15H2,5_H)}$&$Hierarchy$\\
Type &&$(GeV)$&$(GeV)$&$(GeV)$&$(GeV)$&$(GeV)$&$(GeV)$& \\
\hline
\hline
     ${\rm I}$&$37.3$&$3.46\times 10^{15}$&$1.4\times 10^8$&$3\times 10^{12}$&$10^{12}$&$10^{11}$&$3.6\times 10^9$&NO\\
&$37.3$&$3.46\times 10^{15}$&$1.4\times 10^8$&$3\times 10^{12}$&$10^{12}$&$10^{11}$&$3.2\times 10^9$&IO\\
\hline
\hline
II&$37.50$&$3.46\times10^{15}$&$6.3\times 10^8$&$3\times 10^{13}$&$ 10^{13}$&$10^{12}$&$2.4\times 10^{10}$&NO\\
&$37.50$&$3.46\times 10^{15}$&$6.3\times 10^8$&$3\times 10^{13}$&$ 10^{13}$&$10^{12}$&$2.0\times 10^{10}$&IO\\
\hline
\hline  
III&$37.55$&$3.46\times 10^{15}$&$8\times 10^8$&$3\times 10^{13}$&$ 10^{13}$&$5\times 10^{12}$&$2.6\times 10^{11}$&NO\\
&$37.55$&$3.46\times 10^{15}$&$3\times 10^9$&$3\times 10^{13}$&$ 10^{13}$&$5\times 10^{12}$&$2.1\times 10^{11}$&IO\\
\hline
\hline
\end{tabular}
\caption{RG solutions for mass scales and GUT coupling constant in concordance with  precision gauge coupling unification and scalar triplet  masses fitting the neutrino oscillation data and leptogenesis  in the case of the TTM  embedding in non-supersymmetric SU(5) for the fermionic triplet mass $M_{\Sigma}=500$ GeV.}
\label{tab:RGEsol}
\end{table}


It is to be noted that in Table \ref{tab:RGEsol}, each of the  solutions ({\rm I}, II, and III) represents unification of a pair of  corresponding solutions presented in Table 3 and as explained through eq.(\ref{eq:set1}),eq.(\ref{eq:set2}), eq.(\ref{eq:set3}) and eq.(\ref{eq:tri}).
Whereas Sol.I represents unification of 
successful $\tau-$ flavoured leptogenesis, Sol.II and Sol.III
represent successful unification of unflavourd leptogenesis. All the three sets of solutions 
are compatible with successful fits to neutrino oscillation data for both NO and IO types of mass hierarchies.   
  
 In Sec.\ref{sec:fdm} we have discussed  phenomenological importance of triplet fermionic DM covering the mass range $M_{\Sigma}={\cal O}(500-3000)$ GeV all of which are consistent with direct detection cross section limits measured by XENON1T \cite{Aprile:2017,Aprile:2018}.  Out of these masses, the
perturbatively estimated resonance value $M_{\Sigma}=2.4$ TeV (or Sommerfeld enhancement boosted value $M_{\Sigma}\ge 2.74$ TeV) corresponding to dominant fermionic triplet thermal DM has been noted to account for the entire observed cosmological  relic density. 
An example of unification solutions for such heavier $\Sigma$ under class (i) solution category  for  $M_{\Sigma}=2.4$ TeV is presented in Table \ref{tab:RGEsol2}
accommodating all the three sets of bench mark solutions of Table \ref{tab:masspara} as explained further through
eq.(\ref{eq:set1}), eq.(\ref{eq:set2}),  eq.(\ref{eq:set3}), and eq.(\ref{eq:tri}).
Analogous to the $M_{\Sigma}=500$ GeV example of class (ii) solutions applicable for lighter $\Sigma$ presented in Table \ref{tab:RGEsol},  we have thus established one-to-one correspondence of ununified solutions of Table \ref{tab:masspara}  with
the resonance mass $M_{\Sigma}=2.4$ TeV example of our class (i) solutions.  
\\
\begin{table}[h!]
\centering
\begin{tabular}{|c|c|c|c|c|c|c|c|c|}
\hline
\hline
Sol.&$\alpha_G^{-1}$&$M_U$&$M_{C_F}$&$M_{\Delta_1}$&$m_{(15H1,5_H)}$&$M_{\Delta_2}$&$m_{(15H2,5_H)}$&$Hierarchy$\\
Type    &&$(GeV)$&$(GeV)$&$(GeV)$&$(GeV)$&$(GeV)$&$(GeV)$& \\
\hline
\hline
     ${\rm I}$&$38.1$&$2.3\times 10^{15}$&$1.4\times 10^9$&$3\times 10^{12}$&$10^{12}$&$10^{11}$&$3.6\times 10^9$&NO\\
&$38.1$&$2.3\times 10^{15}$&$1.4\times 10^9$&$3\times 10^{12}$&$10^{12}$&$10^{11}$&$3.4\times 10^9$&IO\\
\hline
\hline
II&$38.3$&$2.3\times10^{15}$&$2.7\times 10^9$&$3\times 10^{13}$&$2\times 10^{13}$&$10^{12}$&$2.4\times 10^{10}$&NO\\
&$38.3$&$2.3\times 10^{15}$&$2.7\times 10^9$&$3\times 10^{13}$&$2\times 10^{13}$&$10^{12}$&$2\times 10^{10}$&IO\\
\hline
\hline  
III&$38.35$&$2.3\times 10^{15}$&$3\times 10^9$&$3\times 10^{13}$&$2.6\times 10^{11}$&$5\times 10^{12}$&$2.6\times 10^{11}$&NO\\
&$38.35$&$2.3\times 10^{15}$&$3\times 10^9$&$3\times 10^{13}$&$2.6\times 10^{11}$&$5\times 10^{12}$&$2.1\times 10^{11}$&IO\\
\hline
\hline
\end{tabular}
\caption{RG solutions for mass scales and GUT coupling constant in concordance with  precision gauge coupling unification and scalar triplet  masses fitting the neutrino oscillation data and leptogenesis  in the case of the TTM  embedding in non-supersymmetric SU(5) for the fermionic triplet DM perturbative resonance mass $M_{\Sigma}=2.4$ TeV.}
\label{tab:RGEsol2}
\end{table}

\begin{figure}
\begin{center}
\includegraphics[width=10cm,height=8cm]{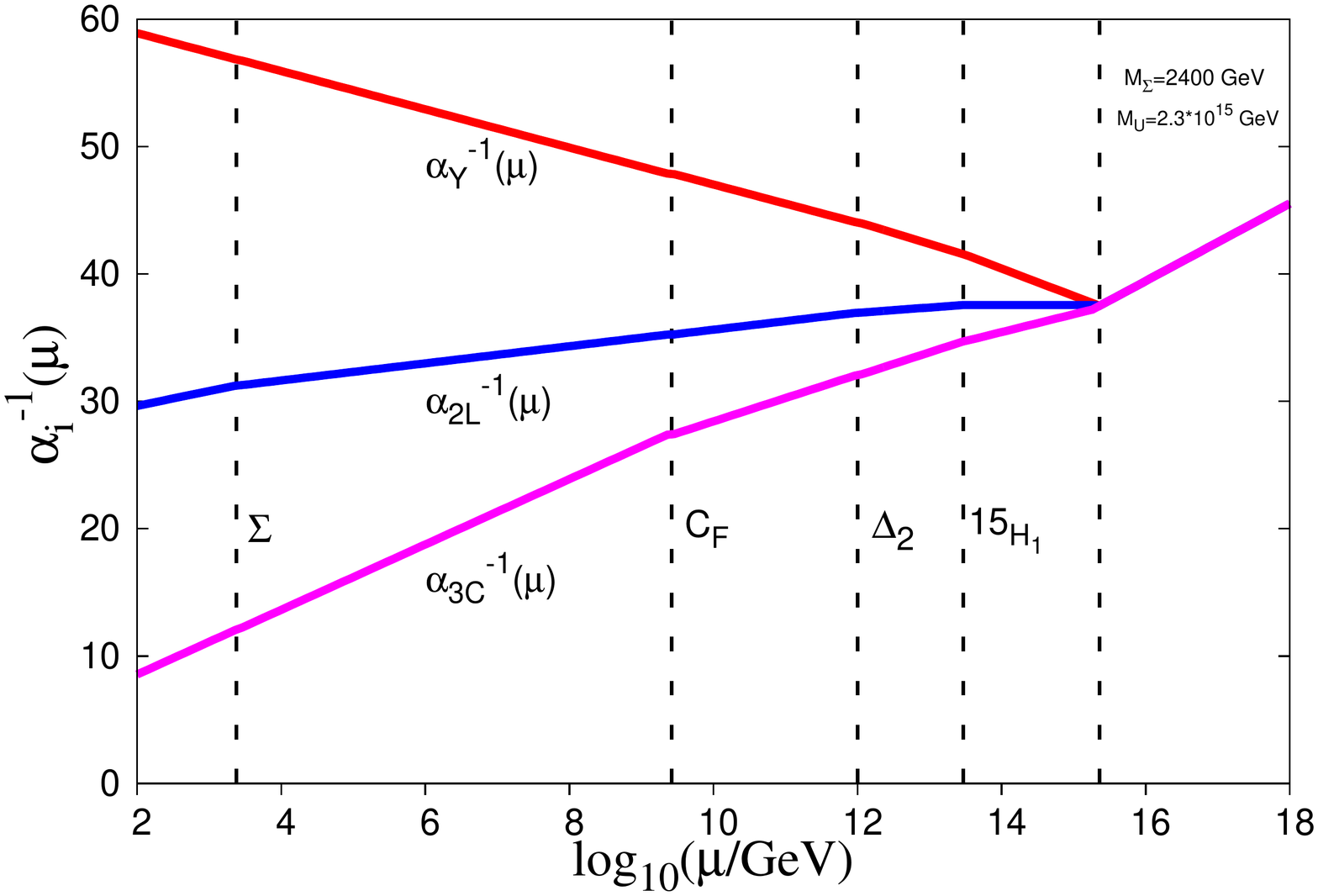}
\caption{RG evlutions  of inverse fine structure constants of the SM gauge group
  $SU(3)_C\times SU(2)_L\times U(1)_Y$  showing unification at the GUT scale $M_U=2.30\times 10^{15}$ GeV in the presence of masses of lighter scalar triplet 
    $M_{\Delta_2}=10^{12}$ GeV,  heavier scalar triplet $M_{\Delta_1}=M_{15_{H1}}=3\times
10^{13}$ GeV, weak triplet fermion   $M_{\Sigma}=2.4$ TeV,  and colour octet fermion $M_{C_F}=2.68\times 10^9$ GeV.}
\label{fig:invHSIG}
\end{center}
\end{figure}

RG evolutions of the three inverse fine structure constants under class (i) solutions with $M_{\Sigma}=2.4$ TeV, $M_{C_F}=3.16\times 10^9$ GeV, $M_{\Delta_2}=10^{12}$ GeV, and $M_{\Delta_1}=3\times 10^{13}$ GeV are presented in Fig. \ref{fig:invHSIG} showing precision gauge coupling unification at $M_U\simeq 2.3\times 10^{15}$ GeV. In the following Sec. \ref{sec:taup} we have noted that such $\simeq 30\%$ decrease in the unification scale, compared to $M_U=3.46\times 10^{15}$ GeV for $M_{\Sigma}=500$ GeV given in Table \ref{tab:RGEsol}, poses no problem for proton life time compliance with Hyper-Kamiokande limit \cite{HyperK}  in this model as the unknown mixing $V_{eff}$ can assume correspondingly smaller values
\cite{Perez:2019-1,Perez:2018-2,Perez-1}.
. 

\subsection{\textbf{Proton lifetime estimation}}\label{sec:taup} 
The ongoing experimental search for proton decay at Hyper-Kamiokande has
set the lower limits on the  life
time for the decay mode $p\to e^+\pi^0$ \cite{HyperK}
\begin{equation}
 \tau_p^{HypK.}(p\to e^+\pi^0)~\ge ~8\times 10^{34}~~{\rm yrs},\label{tauphyp}
\end{equation}
whereas the Super-Kamiokande bound \cite{SuperK} is
 $\tau_p^{SupK.}(p\to e^+\pi^0)~\ge ~1.6\times 10^{34}$ yrs.
We discuss how far the present SU(5) model is capable of respecting these bounds.
 Including strong and electroweak renormalisation 
effects on the ${\rm dim}. 6$ operator and taking into account fermion mixing\cite{Perez:2019-1,Perez:2018-2,Perez-1}, chiral symmetry breaking
effects, and lattice gauge theory  estimations, the decay width  \cite{Buras:1978,Abbot-Wise:1980,Aoki:2007,Aoki:2014,Aoki:2017} is

\beq
\Gamma(p\rightarrow e^+\pi^0)=\frac{\pi m_p \alpha_G^2}{2 M_U^2}\times A^2 V_{eff}^2|{\cal M}|^2 .  \label{eq:width}
\eeq
In eq.(\ref{eq:width}) $m_p=939$ MeV, $A=A_LA_S$, $A_L(A_S)=$ Long (short) distance renormalization factor which has been estimated below, and
 the matrix element ${\cal M}$ is identified as
\beq
{\cal M}= \langle \pi^0|(ud)_R u_L|p\rangle , \label{eq:hadm1}
\eeq 
 leading to the formula  for proton lifetime is  
\begin{eqnarray}
&&\Gamma^{-1}(p\rightarrow e^+\pi^0)=\tau_P(p \to e^+\pi^0) \nonumber\\
&& = \frac{2}{\pi m_p}\frac {M_{U}^4}{\alpha_G^2}\frac{1}{A^2V_{eff}^2} \frac{1}{
|{\cal M}|^2} . \label{eq:taup}
\end{eqnarray} 
 Denoting the generational mixing indices by $\alpha \beta$ in the mixing matrix $V_{i}^{\alpha,\beta}
(i=1,2,3)$, the parameter $V_{eff}$ is defined as \cite{Perez:2019-1,Perez:2018-2,Perez-1}
\beq
V_{eff}^2=\left(|(V_2^{11}+V_{UD}^{11}(V_2V_{UD})^{11})|^2+|V_3^{11}|^2\right). \label{eq:Rmix}
\eeq
Dropping the superscripts for convenience
\begin{eqnarray}
V_2&=& E_C^{\dagger}D,\,\, V_3=D_C^{\dagger}E,\,\,\nonumber\\
 V_3&=& D_C^{\dagger}E,\,\,V_{UD}=U^{\dagger}D. \label{eq:defV}
\end{eqnarray}
Here U, E, and D matrices in eq.(\ref{eq:defV}) are the diagonalising matrices of Yukawa matrices for up-quarks, charged leptons and down quarks, respectively:
\begin{eqnarray}
Y_u^{diag}&=&U_C^TY_uU,\,\, Y_d^{diag}=D_C^TY_d D,\nonumber\\
Y_e^{diag}&=&E_C^TY_eE.          \label{eq:yukdiag}
\end{eqnarray}
Although the elements of the CKM matrix $V_{UD}$ are known, the mixing matrices $V_2$ and $V_3$ are  unknown in such SU(5) models making the parameter $V_{eff}$
undetermined by the model itself.
For our estimation we will use a  wider range of values of this  parameter  :$V_{eff}=0.1 -1.0$ \cite{Perez:2019-1,Perez:2018-2} to have an idea how far present and future measurements can constrain this model compared to similar studies in SO(10) \cite{cps:2019}. Although in the anti-neutrino decay channels $p\to \pi^+ {\bar {\nu}}$ and $p \to K^+{\bar {\nu}}$, it is possible to predict upper bounds on the respective lifetimes \cite{Perez:2019-1,Perez:2018-2}, we plan to carry out these separately including threshold effects which have been shown to play a significant role \cite{psna:2017}.  
Using recent estimations  \cite{Aoki:2014,Aoki:2017}
\begin{equation}
{\cal M}=\langle \pi^0
|(ud)_R u_L|p\rangle = -0.103(23)(34) {\rm GeV}^2,\,\, \label{Aoki:2014}
\end{equation}
\begin{equation} 
{\cal M}= -.131(4)(13){\rm GeV}^2,\,\, \label{Aoki:2017}
\end{equation}
 we will estimate $\tau_P$ for $V_{eff}=0.1-1.0$ \cite{Perez:2019-1} and both these matrix inputs from \cite{Aoki:2014,Aoki:2017}. 
\subsubsection{\textbf{Long and short distance corrections for ${\rm dim.} 6$ operator}}\label{sec:dim6}
\begin{figure}[h!]
\begin{center}
\includegraphics[scale=0.4]{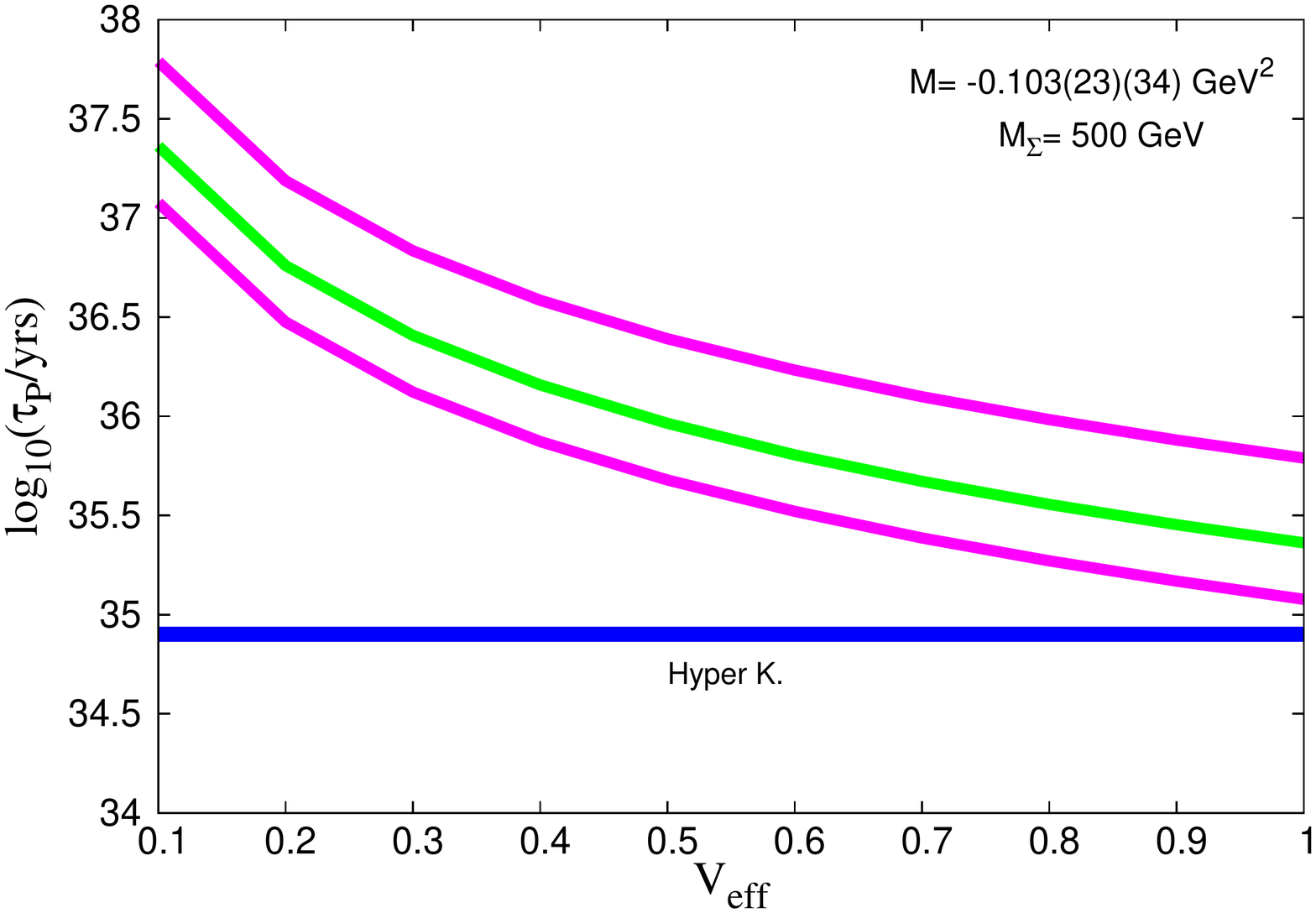}
\end{center}
\begin{center}
\includegraphics[scale=0.4]{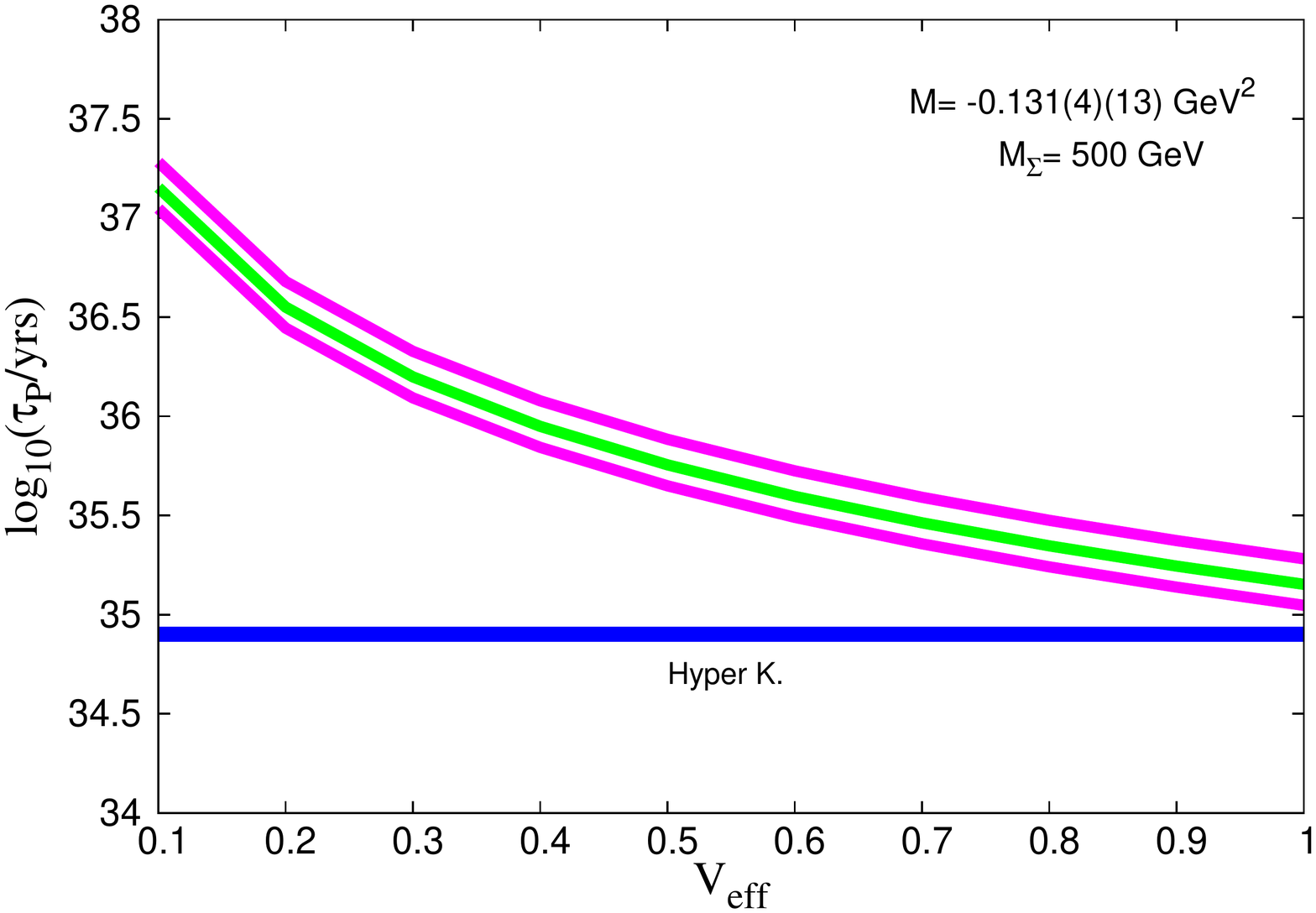}
\end{center}
\caption{Proton lifetime estimation  for solutions of Table \ref{tab:RGEsol}
corresponding to $M_{\Sigma}=500$ GeV
 as a function of the unknown mixing parameter $V_{eff}$ for the decay mode $p\to e^+\pi^0$ for two values of the matrix element
(i) ${\cal M}= -0.103(23)(34)$ GeV$^2$ (upper figure), (ii)  ${\cal M} = -0.131 (4)(13)$ GeV$^2$ (lower figure). In each figure the upper and lower curves represent respective uncertainties in the matrix element \cite{Aoki:2014,Aoki:2017}. The  horizontal lines in both the figures indicate the Hyper-Kamiokande limit \cite{HyperK}:\,\,$\tau_P=8\times 10^{34}$ yrs.} 
\label{fig:taup5}
\end{figure}

The proton lifetime formula discussed in Sec.\ref{sec:taup} contains
short distance ($A_{S}$) and long distance ($A_{L}$) renormalisation
factors.
 The long distance renormalisation factor $A_L$ occuring in the proton decay
 width of eq.(\ref{eq:width}) and in eq.(\ref{eq:taup}) of Sec.\ref{sec:taup}) represents evolution
 of the dim.6 operator for mass scales $\mu=Q=2.3$ GeV to
 $\mu=M_{Z}$. This factor $A_L$ is the same for all the three models
\beq
A_L=\left(\frac{\alpha_3(M_b)}{\alpha_3(M_{Z})}\right)^{6
  \over 23}\left(\frac{\alpha_3(Q)}{\alpha_3(M_{b})}\right)^{6 \over
  25}. \label{eq:AL} 
\eeq
Numerically we estimate its value to be
\beq
A_L=1.15  
\eeq
The  short distance renormalisation factor is  
\begin{eqnarray}
A_{S}&=&\left(\frac{\alpha_3(M_Z)}{\alpha_3(M_{\sigma_F})}\right)^{-2 \over a^{(1)}_3}\left(\frac{\alpha_3(M_{\sigma_F})}{\alpha_3(M_{C_F})}\right)^{-2 \over a^{(2)}_3}\left(\frac{\alpha_3(M_{C_F})}{\alpha_3(M_{\Delta_2})}\right)^{-2 \over a_3^{(3)}} \nonumber\\
&\times& \left(\frac{\alpha_3(\Delta_2)}{\alpha_3(M_{\Delta_1})}\right)^{-2 \over a^{(4)}_3}\left(\frac{\alpha_3(M_{\Delta_1})}{\alpha_3(M_{U})}\right)^{-2 \over a^{(5)}_3}\nonumber\\
&\times& \left(\frac{\alpha_2(M_Z)}{\alpha_2(M_{\sigma_F})}\right)^{-9 \over {4a^{(1)}_2}}\left(\frac{\alpha_2(M_{\sigma_F})}{\alpha_2(M_{C_F})}\right)^{-9 \over {4a^{(2)}_2}}\left(\frac{\alpha_2(M_{C_F})}{\alpha_2(M_{\Delta_2})}\right)^{-9 \over {4a^{(3)}_2}} \nonumber\\
&\times& \left(\frac{\alpha_1(M_Z)}{\alpha_1(M_{\sigma_F})}\right)^{-11 \over {20a^{(1)}_1}}\left(\frac{\alpha_1(M_{\sigma_F})}{\alpha_1(M_{C_F})}\right)^{-11\over {20a^{(2)}_1}}\left(\frac{\alpha_1(M_{C_F})}{\alpha_1(M_{\Delta_2})}\right)^{-11 \over {20a^{(3)}_1}} \nonumber\\
&\times& \left(\frac{\alpha_1(M_{\Delta_2})}{\alpha_1(M_{\Delta_1})}\right)^{-11 \over {20a^{(4)}_1}}\left(\frac{\alpha_1(M_{\Delta_1})}{\alpha_1(M_{U})}\right)^{-11\over {20a^{(5)}_1}},
\label{eq:SR}
\end{eqnarray}
where $M_{\Delta_1}=M_{{15}_{H1}}$. Using the one-loop beta function coefficients and coupling constant values at the respective mass scales $\mu= M_Z, M_{\Sigma}, M_{C_F}, M_{\Delta_2}, M_{\Delta_1},M_U$, we have estimated
\beq
A_{S}=2.238,\, A=A_{S}A_{L}=2.576\,\, \label{eq:srsl}
\eeq 
Using these model parameters in eq.(\ref{eq:taup}), proton lifetime estimations for $M_U=10^{15.54}$ GeV   are
shown in Fig.\ref{fig:taup5} for the $p\to e^+\pi^0$ decay mode as a function of the parameter $V_{eff}=0.1-1.0$.
Out of different experimental bounds \cite{SuperK,HyperK}, the stronger one is due to \cite{HyperK}.
It is clear that the values of $V_{eff}\ge 2$ are ruled out by both the current  limits \cite{SuperK,HyperK}  unless threshold effects are taken into account. 
But a large region of the parameter space for $V_{eff} < 1$ is likely to remain unconstrained in near future unless the actual proton decay is detected to fix this model parameter. Future improved measurements over the current  Hyper-Kamiokande limit \cite{HyperK} would further constrain the parameter space spanned by $V_{eff}$ \cite{Perez:2019-1,Perez-1}.\\   

\begin{figure}[h!]
\begin{center}
\includegraphics[scale=0.4]{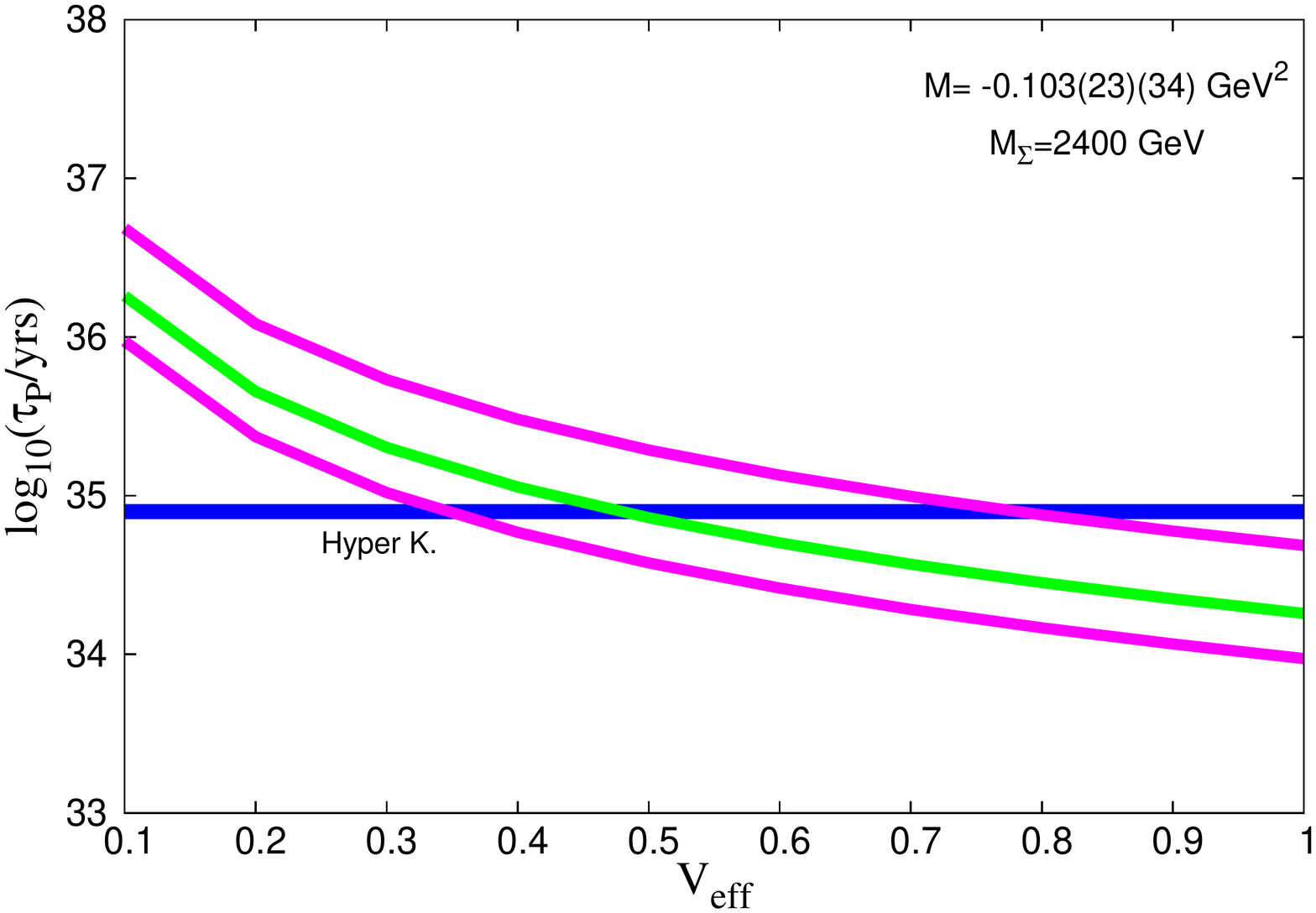}
\end{center}
\begin{center}
\includegraphics[scale=0.4]{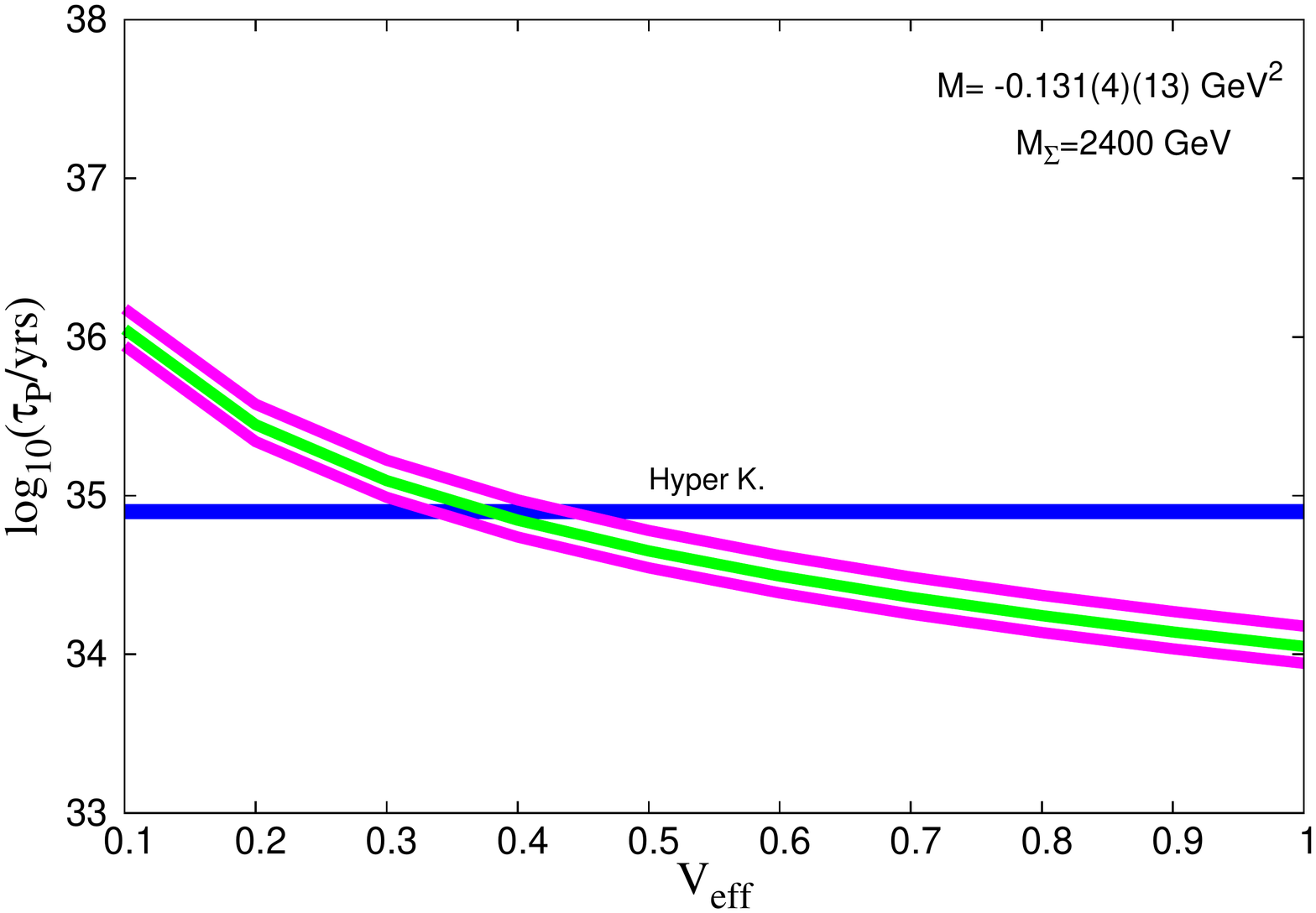}
\end{center}
\caption{Proton lifetime estimation  for solutions of Table \ref{tab:RGEsol2}
corresponding to $M_{\Sigma}=2400$ GeV
 as a function of the unknown mixing parameter $V_{eff}$ for the decay mode $p\to e^+\pi^0$ for two values of the matrix element
(i) ${\cal M}= -0.103(23)(34)$ GeV$^2$ (upper figure), (ii)  ${\cal M} = -0.131 (4)(13)$ GeV$^2$ (lower figure). In each figure the upper and lower curves represent respective uncertainties in the matrix element \cite{Aoki:2014,Aoki:2017}. The  horizontal lines in both the figures indicate the Hyper-Kamiokande limit \cite{HyperK}:\,\,$\tau_P=8\times 10^{34}$ yrs.} 
\label{fig:taupHSig}
\end{figure}

Compared to $M_{\Sigma}=500$ GeV solutions of Table \ref{tab:RGEsol} for which the GUT scale is $M_U=3.46\times 10^{15}$ GeV, our solutions for heavier fermionic triplet mass $M_{\Sigma}=2.4$ TeV yields $M_U=2.3 \times 10^{15}$ GeV which is nealy $30\%$ lighter. Proton lifetime estimations  corresponding  to $M_{\Sigma}=2.4$ TeV solution are presented in the upper and lower panels of Fig.\ref{fig:taupHSig} for lattice gauge theory  matrix elements  ${\cal M}=-0.103$ GeV$^2$ and ${\cal M}=-0.131$ GeV$^2$, respectively \cite{Aoki:2014,Aoki:2017}.    
Whereas Hyper-Kamiokande experimental limit could be satisfied with the unknown mixing parameter $V_{eff} \simeq 1$ for $M_{\Sigma}=500$ GeV solutions, lower values of $V_{eff}=0.4-0.8$ are needed for the $M_{\Sigma}=2.4$ TeV solutions. For the values of the entire range of triplet fermion masses used as input parameter $M_{\Sigma} \simeq {\cal O}(500-3000)$ GeV, while maintaining the same values of $\Delta_1$ and $\Delta_2$ masses but with  variations over the corresponding 
values of $M_{C_F}$ and $M_U$  given in Table \ref{tab:RGEsol} and Table \ref{tab:RGEsol2}, we  
find precision coupling unification yielding  $M_U\simeq (10^{15.53}-10^{15.28})$ GeV all of which are capable of yielding proton lifetimes longer than Hyper-Kamiokande limit by suitably adjusting the  mixing parameter $V_{eff}$. Thus the entire range of masses $M_{\Sigma}\simeq {\cal O}(500-3000)$ GeV are consistent with precision gauge coupling unification in the present SU(5) model predicting the same sets of  values of heavy $\Delta_1, \Delta_2$ masses fitting the neutrino oscillation data and baryon asymmetry of the Universe through similar ansatz for unflavoured or partially flavoured leptogenesis.     

\section{Fermionic triplet dark matter}\label{sec:fdm}
In Sec.\ref{sec:uni} we have found that the embedding of the TTM in SU(5) with precision gauge coupling unification predicts a triplet fermion $\Sigma (3,0,1)$  and a colour octet fermion $C_F(1,0,8)$ both of which have been  suggested to originate from the non-standard fermionic representation ${24}_F \subset $ SU(5). 
The phenomenology of a hyperchargeless triplet fermionic DM in the
non-SUSY model is similar
 to that of the wino DM in MSSM and SUSY
GUTs which has been
 investigated recently  \cite{Cirelli:2006, Cirelli:2007,Frig-Ham:2010,Hryzcuk:2014,Cirelli:2014,Dutta:2013,Hisano:2007},
and also continues to be a subject of current importance
\cite{Hryzcuk:2014,Cirelli:2014,Dutta:2013}. It is pertinent to note that if  $\Sigma$ 
represents a non-thermal DM, it can account for the entire value of the observed relic density \cite{Planck15,wmap} for each value  of its hitherto experimentally  undetermined mass  $M_{\Sigma}
\,\,\langle\,\, 3000$ GeV \cite{Hryzcuk:2014,Cirelli:2014,Dutta:2013} ,  which have been found to complete precision gauge coupling unification through SU(5) as noted above.
The wino dominated DM cosmological relic density prediction has been also normalised to its observed value
even at $M_{\Sigma}\simeq 100$ GeV   while searching for its signature at LHC with proposed upgradation of beam luminosity \cite{Dutta:2013}. However, our discussions in this work are mainly related to thermal DM. \\      

 In the present SU(5) model $\Sigma$ is  stabilised by an additional $Z_2$ discrete symmetry. In Table \ref{tab:su5z2} we have presented the $Z_2$ charges of all particles and representations of the present SU(5) theory where the $Z_2=+1$ for ${24}_F,\Sigma, C_F, 5_H$ but both ${\bar 5}_F$ and ${10}_F$ have $Z_2=-1$. 

\begin{table}[!h]
\caption{Particle representations of unified two-triplet model for neutrino mass, baryon asymmetry and dark matter with respective charges under $G_{213}\times Z_2$ and $SU(5) \times Z_2$. 
The  second and the third generation fermions not shown in this Table have identical transformation properties as those of the first generation. The scalar singlet DM $\xi$ compensates for the relic density missed by the fermionic DM $\Sigma$ constrained by indirect experimental searches and also satisfies direct detection experimental bounds  discussed in Sec. \ref{sec:fxidm}.}   
\label{tab:su5z2}
\begin{center}
\begin{tabular}{|c|c|c|c|}
\hline 
{ Particle } & $G_{213}$ & $SU(5)$& $Z_2$\\
\hline
$(\nu, e)^T_L$ & $(2,-1/2,1)$ &${\bar 5}_F$& $-1$\\
$e_R$ & $(1,-1,1)$ &${10}_F$& $-1$\\
$(u,d)_L^T$ & $(2,1/6,3)$ &${10}_F$& $-1$\\
$u_R$ & $(1, 2/3,3)$ &${10}_F$& $-1$\\
$d_R$ & $(1,-1/3,3)$ &${\bar 5}_F$& $-1$\\
$\Sigma$&$(3,1,0)$&${24}_F$&$+1$\\
$C_F$&$(1,0,8)$&${24}_F$&$+1$\\
\hline
$\phi$ & $(2,1/2,1)$ &${5}_H$ &$+1$\\
$\Delta_1$ & $(3,-1,1)$ &${15}_{H1}$&$+1$\\ 
$\Delta_2$ & $(3,-1,1)$ &${15}_{H2}$&$+1$\\
$S_{24}$&$(1,0,1)$&${24}_H$&$+1$\\
\hline
$\xi$&$(1,0,1)$&$1_H$&$-1$\\
\hline
\hline
 \end{tabular}
\end{center}
\end{table}

These assignments rule out type-III seesaw mediation  \cite{Bajc-gs:2007} by the fermion triplet $\Sigma $ or type-I seesaw  mediation  by the fermion singlet $S_F(1,0,1)\subset {24}_F$ establishing type-II seesaw as the only allowed source of neutrino mass in such two-triplet embedding  through $SU(5)\times Z_2$. These $Z_2$ charge assignments further forbid loop mediation through vertex correction diagrams of Higgs scalar  triplets  bilepton decays (i,e $\Delta_{1,2}\to ll$)  by the singlet or triplet components of ${24}_F$ which further establishes the hypothesis that  the two triplets are rudimentary components for baryon asymmetry generation through leptogenesis.\\
 The even value of the global discrete charge ($Z_2=+1$) of fermion triplet DM $\Sigma(1,3,0)\subset {24}_F$, compared to odd (even) discrete charge of standard fermion (Higgs scalar), guarantees stability
of the DM by ruling out Yukawa interactions with SM fermions. Also the triplet fermion decay through  
discrete symmetry permitted couplings ${24}_F^2{24}_H$  is kinematically forbidden as all other relevant masses are far more heavier than $M_{\Sigma}$. \\
\subsubsection{\textbf{Relic density constraints}}\label{sec:frelic}
The only interaction of the
DM fermion with standard model particles is through gauge interaction
that leads to the well known mass difference \cite{Cirelli:2006}
$M_{\Sigma^{\pm}}-M_{\Sigma^0}=166\, {\rm  MeV}$ 
where we have denoted the mass of the charged (neutral) component of $\Sigma$ as
$M_{\Sigma^{\pm}}(M_{\Sigma^0})$.\\   
We now discuss the implications of our unification solution $M_{\Sigma}\sim {\cal O}(1)$ TeV as WIMP dark matter \cite{Cirelli:2006,Cirelli:2007,Cirelli:2008,Hryzcuk:2014,Cirelli:2014}. Defining  $\Gamma(H)$ as the particle decay rate (Hubble parameter), at a certain stage of evolution
of the Universe a particle species is said to be coupled if $\Gamma > H$ or decoupled if $\Gamma < H$. The
WIMP DM particle has been decoupled from the thermal bath at some early epoch and has remained as a thermal relic.
For estimation of DM relic density \cite{Cirelli:2006,Cirelli:2007,Ma-Suematsu:2008}  the corresponding  Boltzmann
equation \cite{Kolb:1990,Bertone:2004,Gondolo:1990dk,Hisano:2007} is solved approximately
\beq
\frac{dn}{dt}+ 3Hn =- \langle \sigma v \rangle (n^2- n_{eq}^2), \label{eq:BEsig}    
\eeq
where $n =$ actual number density at a certain instant of time (denoted as t in eq.(\ref{eq:BEsig})), $n_{eq} =$ equilibrium number density of  DM particle, $ v =$ velocity, and $\langle \sigma v\rangle =$ thermally averaged annihilation
cross section. Approximate solution of Boltzmann equation gives the expression of relic density 
\beq
{\langle \Omega h^2 \rangle}_{\Sigma} =\frac{1.07 \times 10^9 x^{\Sigma}_F}{\sqrt {g_*} M_{pl}\langle \sigma^{\Sigma}_{eff}|v| \rangle}, \label{eq:omh2}
\eeq
where $x^{\Sigma}_F = M_{\Sigma} /T_F $, $T_F =$ freezeout temperature, $g_{*} =$ effective number of massless degrees of
freedom and $M_{pl} = 1.22 \times 10^{19}$ GeV, and the subscript or superscript
indicate respective quantities associated with the fermion triplet DM $\Sigma(3,0,1)\equiv \Sigma$. 
 The value of $x^{\Sigma}_F$ is computed through  iterative solution of the equation 
\beq
 x^{\Sigma}_F = ln \left[\frac{1}{2\pi^3}\sqrt {\frac{45}{2}}\frac{M_{pl} M_{\Sigma}\langle\sigma^{\Sigma}_{eff} |v|\rangle}{\sqrt {g_*x^{\Sigma}_F}}\right].  \label{eq:xf}
\eeq  
The relic abundance of the neutral component $\Sigma^0$ is theoretically  
estimated  by taking into account the annihilations and co-annihilations of $\Sigma^0$ itself and $\Sigma^+\Sigma^-$. All such relevant cross sections, $\sigma(\Sigma^0,\Sigma^0),\sigma(\Sigma^0,\Sigma^{\pm}), \sigma(\Sigma^+,\Sigma^-)$, and $\sigma (\Sigma^{\pm},\Sigma^{\pm})$ are computed. Adding them up with respective weightage factors \cite{Ma-Suematsu:2008}  gives the effective cross section 
$\langle \sigma^{\Sigma}_{eff}|v|\rangle$  which occurs in eq.(\ref{eq:omh2}) and
eq.(\ref{eq:xf}). This effective cross section is used in the right-hand side of eq.(\ref{eq:xf}) while solving it iteratively to determine $x_F^{\Sigma}$ as discussed above.\\
Neglecting mass difference between $\Sigma^{\pm}$ and $\Sigma^0$, it has been shown from perturbative estimation that \cite{Cirelli:2006,Cirelli:2007,Frig-Ham:2010}
\beq
\langle \sigma^{\Sigma} |v|\rangle = \frac{37 g_{2L}^4}{96 \pi M_{\Sigma}^2} \label{anncross}   
\eeq
The correct amount of relic abundance  $\Omega h^2=0.1172-0.1224$ \cite{Planck15,wmap}, is generated for $M_{\Sigma}\sim 2.4$ TeV \cite{Ma-Suematsu:2008}. This mass value  emerges from perturbative estimations of annihilation and co-annihilation cross sections. However, when non-perturbative contributions with Sommerfeld enhancement effect  are included, matching the relic density within $3\sigma$ uncertainty of relic density ($\Omega h^2=0.095-0.125$) needs $M_{\Sigma}\sim 2.75 \pm 0.15$ TeV \cite{Cirelli:2006,Cirelli:2007,Frig-Ham:2010,Hisano:2007}. Heavier masses $M_{\Sigma}=3.0-3.2$ TeV \cite{Cirelli:2014,Hryzcuk:2014} and $M_{\Sigma}\simeq 4$ TeV \cite{Mohanty:2012} have been also suggested.
Even though $\Sigma$ with a sub-dominant thermal DM mass $M_{\Sigma} < 3$ TeV  below the Sommerfeld enhanced resonance value \cite{Hryzcuk:2014,Cirelli:2014} can not account for the observed  cosmological relic density, it can succeed in doing so if it is  non-thermal \cite{Hryzcuk:2014,Cirelli:2014,Dutta:2013}.

\subsubsection{\textbf{Direct detection and collider searches}}\label{sec:dircoll}
\par\noindent{\underline{(a). Direct detection prospects}}\\
In general for elastic scattering of a DM  particle (which is
electrically neutral) off nucleons, either a
standard Higgs or a $Z$-boson exchange is needed in the t-channel of the dominant tree
diagrams. In the absence of such couplings of $\Sigma^0$, a sub-dominant
process occurs by the exchange of two virtual $W^{\pm}$ bosons through a
box diagram \cite{Cirelli:2006}. This process leads to suppression
of spin independent cross section by $2-3$ orders below the
experimentally detectable value. However, such predicted cross sections
are measurable with improvement of detector sensitivities
\cite{Aprile:2017,Aprile:2018}. Currently XENON1T collaboration \cite{Aprile:2018} has  measured  the lowest value of this cross section up to $4.1\times 10^{-47}$ cm$^2$ which, however,  has been reached only for a lower value of DM mass $M_{DM}=30$ GeV. For heavier DM masses the measured value of direct detection cross   section increases.

The inelastic scattering off nucleons with a charged component
($\Sigma^+$ or $\Sigma^-$) is prevented because of kinematic
constraints since the mass difference, $m_{\Sigma^+}-m_{\Sigma^0}=166
$ MeV, is about three orders of magnitude above the kinetic energy of $\Sigma$
and also much above the proton-neutron mass difference, $m_n-m_p\sim
2$ MeV.

\begin{figure}
\begin{center}
\includegraphics[width=8cm,height=8cm]{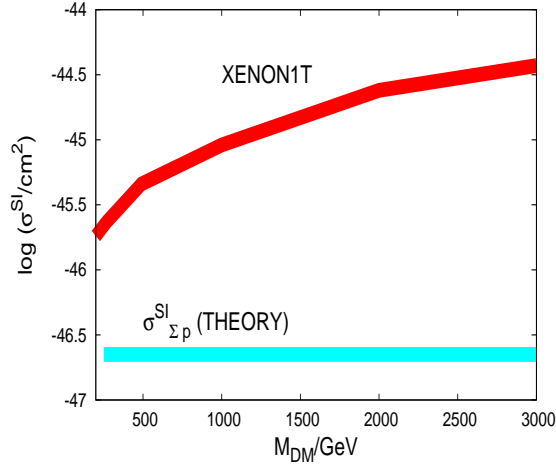}
\caption{Comparison of experimentally determined direct detection cross section  bounds  on spin-independent (SI) DM-Nucleon elastic scattering from XENON1T Collaboration \cite{Aprile:2017,Aprile:2018} presented as red curve with theoretical predictions \cite{Hisano:2015} for $\Sigma p$ elastic scattering denoted as green 
 curve in the triplet fermionic DM mass range $M_{\Sigma}=(270-3000)$ GeV as discussed in the text.}               
\label{fig:xesig}
\end{center}
\end{figure}

Theoretically, for $M_{\Sigma}=(270-3000)$ GeV, the  spin independent $\Sigma p$ elastic scattering cross section
has been recently estimated   for wino DM in SUSY (with decoupled superpartners) and also in non-SUSY SM extension with $\Sigma$ as a minimal DM candidate including QCD effects through two-loop contributions leading to similar results in both cases \cite{Hisano:2015}
\beq
\sigma^{SI}(\Sigma p \to \Sigma p)\simeq \left (2.3^{+.2+.5}_{-.3-.4}\right)\times 10^{-47} cm^2 \label{eq:hsig}
\eeq   
Here the underlying uncertainty due to perturbation theory (input parameters) has been represented by the second (third) term in the right-hand side
of eq.(\ref{eq:hsig}). An interesting new point of this theoretical prediction 
  is that the estimated cross section is clearly above the neutrino background \cite{Billard:2014} which makes 
the future direct detection prospects  for $\Sigma$ more promising.
In Fig.\ref{fig:xesig} we have compared the theoretical prediction of eq.(\ref{eq:hsig}) (presented as green curve) with the experimentally determined direct detection cross section bounds from    
 XENON1T Collaboration \cite{Aprile:2017,Aprile:2018}  in the same DM mass range $M_{\Sigma}=270-3000$ GeV (presented as red curve). Direct detection bounds from LUX-2016 \cite{Akerib:2016}
and Panda X-II \cite{Cui:2017} (not shown in Fig. \ref{fig:xesig}) are either larger or  equal to XENON1T \cite{Aprile:2017,Aprile:2018} data presented by the 
red curve.
As shown in Fig. \ref{fig:xesig}, the experimentally measured  direct detection cross sections increase by $1-3$ orders beyond the theoretical prediction in the triplet fermionic DM  mass range $M_{\Sigma}={\cal O}(500-3000)$ GeV \cite{Hisano:2015}.

It is clear that for the mass range $M_{\Sigma}\sim 2-3$ TeV, the spin independent theoretical cross section due to $\Sigma-p$ elastic scattering is about ${\cal O}(10^{-2}-10^{-3})$ times smaller than  the current  XENON1T \cite{Aprile:2017,Aprile:2018} and other direct detection experimental bounds from  LUX-2016 \cite{Akerib:2016} and Panda X-II \cite{Cui:2017} whose measured cross sections are larger than the XENON1T collaboration \cite{Aprile:2018}. But   Fig.\ref{fig:xesig}  also indicates near future 
possibility  of direct experimental detection provided the fermionic triplet DM is in the lower mass region $270\le M_{\Sigma}/{\rm GeV} < 1000$ for which accuracies of  XENON1T cross section measurements have to be improved by 1-2 orders. 


\par\noindent{\underline{(b). Collider signaure}}\\
Prospects of observing signatures of the triplet fermion DM  at colliders  have
been investigated in \cite{Franceschini:2008,Aguilla:2009,Arhrib:2009,Roeck:2009,Cirelli:2014,Dutta:2013}. For $M_{\Sigma}\sim
2.7$ TeV and integrated luminosity of $100$fb$^{-1}$, the DM pair
production cross section at LHC in the channel $pp\to \Sigma\Sigma X$
has been shown to result in only one event \cite{Franceschini:2008,Aguilla:2009,Arhrib:2009}. For better
detection capabilities, upgradation of LHC with twice the current energy and more
luminosity has been suggested \cite{Roeck:2009}.\\
Besides ref.\cite{Arhrib:2009,Roeck:2009},  collider signatures in models where $\Sigma$ mediates type-III seesaw mechanism (which is ruled out in this work by $Z_2$ charge assignments) have been also discussed earlier  \cite{Bajc-gs:2007,Keung-gs:1983} and more recently in \cite{Arindam:2020}.\\     

For detection at $e^+e^-$
collider that requires a collision energy of at least twice the DM mass, observation of
$\Sigma^+\Sigma^-$ pair production is predicted via $Z$ boson
exchange \cite{Cirelli:2006,Franceschini:2008}. The neutral pair $\Sigma^0\Sigma^{0*}$ can be
also produced at LHC (or  $e^+e^-$ collider), although at a suppressed rate, through  one-loop box
diagram mediated by two virtual $W$ bosons.  
After production, such charged  components  would provide a clean signal as they would
manifest in long lived charged tracks due to their decays via
standard gauge boson interactions,  $\Sigma^{\pm}\to W^{\pm} \to
\Sigma^0\pi^{\pm}$, or $\Sigma^{\pm}\to W^{\pm} \to
\Sigma^0 l^{\pm}\nu_l (l=e,\mu)$.
 The production of $e^{\pm}$ and
$\mu^{\pm}$ charged leptons but the absence of $\tau^{\pm}$ due to
kinematical constraint may be another distinguishing experimental
signature of the triplet fermionic DM at LHC. The decay length of associated
displaced vertices is clearly predicted \cite{Cirelli:2006,Franceschini:2008} to be
$L_{\Sigma^{\pm}}\simeq 5.5$ cm. \\

It has been also noted in the context of 
SO(10)  \cite{Frig-Ham:2010} that
the decay product $\Sigma^0$  is stable because of its matter
parity which survives the GUT breaking as a gauged discrete symmetry. As such the production of this neutral component of the
triplet fermion DM
originating from SO(10) will be signalled through missing energy
\cite{Frig-Ham:2010}. This stability feature of  $\Sigma^0$ with its
TeV scale mass has
negligible impact on electroweak precision variables. 
These interesting features are applicable also in the
present model where the assumed $Z_2$ symmetry executes a role analogous to matter parity \cite{Frig-Ham:2010} in SO(10).\\

 LHC at 14 TeV energy and luminosity $1000$ fb$^{-1}$ has been shown to be capable of probing wino DM up to $M_{\Sigma}\simeq 600$ GeV  through vector boson fusion process where the DM ($\equiv$ non-thermal) relic density has been normalised to its observed value at as low a mass as $M_{\Sigma}\simeq 100$ GeV \cite{Dutta:2013}.
Detection possibility of $\Sigma$ in the multi TeV range at high luminosity LHC and future 100 TeV pp collider has been also investigated \cite{Cirelli:2014} in different channels such as: (i) monojet, (ii)monophoton, (iii)vector boson (VB) fusion, and (iv) disappearing tracks. At 100 TeV, the disappearing track channels are likely to probe the resonance mass $M_{\Sigma}=3 $ TeV, deduced including higher order corrections and Sommerfeld enhancement,  which is relevant for thermally produced DM that accounts for whole of observed relic density \cite{Hryzcuk:2014}.
Although $\Sigma$ as sub-dominant thermal DM can not account for observed values of relic density for lighter masses, it has been noted that it can do so for every value of the mass $M_{\Sigma} < 3$ TeV if it is  non-thermal \cite{Hryzcuk:2014,Cirelli:2014,Dutta:2013} whereas the relic density at $M_{\Sigma}\simeq 3-3.2$ TeV can be realised as thermal DM through Somerfeld boosted annihilation cross section derived including higher order contributions \cite{Hryzcuk:2014}.
 Searches in other channels have been found to extend upto $M_{\Sigma}=1.3 (1.7)$ TeV for 3(30) ab$^{-1}$ of integrated luminosity provided systematics are under control \cite{Cirelli:2014}. These masses being away from resonance values, although a thermal $\Sigma$ can not account 
for the entire value of observed relic density, it can do so as a non-thermal DM .  
 For  $M_{\Sigma} > 3.2$ TeV the DM $\Sigma$ can not be treated to be thermally produced as in that case it overcloses the Universe. For such heavy masses nonthermal origin of $\Sigma$ is preferred \cite{Hryzcuk:2014}.        
\subsubsection{\textbf{Prospects from indirect searches}}\label{sec:sigind}
PAMELA \cite{pamela} and FERMI/LAT \cite{fermi} experiments have measured the positron excess akin to 
wino  DM model  which is again confirmed by recent AMS-02 \cite{ams} data. The electron and positron fluxes are still significant 
in the measurement of FERMI/LAT \cite{fermi}.
There are various constraints on the wino dark matter from different search channels such as 
antiprotons, leptons, dark matter halo from diffuse galactic gamma rays, 
 high latitude gamma-ray spectra, galaxy clusters, dwarf spheroids, 
 gamma-ray line feature, neutrinos from the galactic halo, CMB constraints, and
antideuterons \cite{Hryzcuk:2014}. 
In the case of the antiproton search channel, the wino dark matter having mass close to the perturbatively estimated resonance value i.e,
$M_{\Sigma}\simeq 2.4$ TeV, and thin zone of diffusion is  consistent with the antiproton measurement. But such a
 wino dark matter having mass near the resonance, produces very small amount of leptons and large amount of positrons at
very low energy scale. This DM can not solve cosmic ray (CR) lepton puzzle because the lepton data can
rule out the very proximity of resonance.  The galactic $\gamma$ rays also impose a stringent limit on the 
wino DM model. With the inclusion of the $\gamma$ ray constraint, the limit on the wino DM mass changes. If the mass of  DM
 is $M_{\Sigma}\sim 2.5$ TeV and it is in a thin diffusion zone, then it is excluded by the $\gamma$ ray data for a wide variation
of galactic cosmic ray  propagation. There is also a very significant limit on the wino dark matter mass from high latitude
$\gamma$ ray spectra. For a $2.5$ TeV wino DM,
the expected 10 year cross section is $1.5\times 10^{-25}$ $cm^3s^{-1}$ including DM substructures \cite{Hryzcuk:2014}. 
Possible signatures of DM annihilations are given from $\gamma$ ray observations\cite{Han,Hektor} towards nearby galaxy clusters 
 but observations in ref.\cite{HESS,MAGIC,Ajello,Dugger,Zimmer,Huang:12} have not seen any significant limits from $\gamma$ ray excess. The wino dark matter having mass
$M_{\Sigma}\simeq 2.4$ TeV can be ruled out in this search channel whereas all the other masses are allowed in the dwarf spheroids channel \cite{Hryzcuk:2014}. 
 The winos with masses heavier than 2 TeV are excluded by the HESS\cite{HESS} data
at $95\%$ CL. A new method to search for the indirect signals of DM annihilation is obtained due to
the motion of high energy neutrons towards the galactic center. Wino model signals corresponding to
 $M_{\Sigma}\simeq 2.4$ TeV can be observed in this search channel \cite{Hryzcuk:2014}. 
There is also a constraint on the wino dark matter due to the CMB temperature and polarization power
spectra. Taking WMAP-5 \cite{Komatsu} data and with $98\%$ CL,
the DM masses  in the region $2.3$ TeV to $2.4$ TeV have been excluded.
Similarly WMAP-9 \cite{Hinshaw}  excludes the region $2.25-2.46$ TeV. But
the combined search of WMAP-9 with ACT \cite{Sievers,Story} excludes the mass range of $2.18-2.5$
TeV . To search for the dark matter, the most
effective  channel is through antideuterons. 
Due to the smaller signal to back ground ratio at mass $2.5$ TeV, the resultant signal is very low with high uncertainty.
With the theoretical and experimental progress, there may be stringent limit on the wino dark matter \cite{Hryzcuk:2014}.\\ 
In the indirect detection experiment, signals are produced in the annihilation process: DM DM $\rightarrow$ SM particles. Here we have tree level annihilation only to $W^+W^-$ channel. The annihilation cross section to
$W^+W^-$ shows a peak  near $M_\Sigma \sim 2.7$ TeV when 
non-perturbative effect due to Sommerfeld enhancement is taken into account. But this high value of annihilation cross section exceeds the upper limit given by the 
combined analysis  of Fermi/LAT \cite{fermi} and MAGIC \cite{MAGIC}. Thus when the WIMP dark matter is composed of only $\Sigma^0$, it may be somewhat difficult to satisfy  all the  constraints from relic density, and direct and
indirect detection experiments, simultaneously.
 A better agreement with PAMELA \cite{pamela} positron and antiproton fluxes, and the   observed   cosmological relic density \cite{wmap} has been  suggested for a heavier wino DM of mass $M_{\Sigma}\simeq 4$ TeV \cite{Mohanty:2012} in which case precision gauge coupling unification in the present approach is possible.

 Despite the constraints discussed above, the triplet fermion as a minimal, dominant thermal DM appears to be an attractive idea  which is also predicted by the present unification framework. One class ($=$ class (i)) of our  solutions discussed in  Sec.
\ref{sec:su5}) of this SU(5) unification program with vacuum stability of the scalar potential ensured by intermediate mass scalar singlet threshold effect (as discussed below) supports $\Sigma$ as dominant thermal  DM of the Universe with $M_{\Sigma}\simeq 2.4$ TeV, or heavier \cite{Cirelli:2006,Cirelli:2007,Frig-Ham:2010,Cirelli:2014,Hisano:2015,Mohanty:2012}.\\ 
\subsubsection{\textbf{Vacuum stability}}\label{sec:vstab}
It is well known  that the Higgs quartic coupling $\lambda_{\phi}$ of the SM scalar potential 
\begin{equation}
V_{std}=-{\mu_{\phi}}^2 \phi^{\dagger}\phi+ \lambda_{\phi}(\phi^{\dagger}\phi)^2, \label{eq:smpot}
\end{equation}
runs negative  \cite{Espinosa:2012} for SM Higgs field values 
$|\phi| \sim {\cal O}(10^9-10^{10})$ GeV causing instability to the SM vacuum that is associated with the Higgs mass $m_{\phi}\simeq 125$ GeV. In the renormalisation group (RG) predictions discussed in Sec.\ref{sec:su5} the TTM unification has permitted a variety of solutions for neutrino mass and leptogenesis some of which do not have this problem because of the allowed mass scale of the  triplets  which have  threshold corrections on the quartic coupling \cite{pcns:2020}
\bea
\Delta {\lambda_{\phi}}^{(2)}&=&\frac{\mu_2^2}{M_{\Delta_2}^2}\Theta(|\phi|-M_{\Delta_2}),     \nonumber\\ 
\Delta {\lambda_{\phi}}^{(1)}&=&\frac{\mu_1^2}{M_{\Delta_1}^2}\Theta(|\phi|-M_{\Delta_1}). \label{eq:thlam}
\eea
where $\Theta(x)=$  Heavyside function. Even if the effect due to heavier $\Delta_1$ is neglected,  positive corrections due to lighter $\Delta_2$ removes vacuum instability if $M_{\Delta_2} \simeq {\cal O}(10^{10})$ GeV \cite{pcns:2020}. In fact we have a class of solutions belonging to  partially flavoured or $\tau-$ flavoured leptogenesis which does have such intermediate mass solutions for $M_{\Delta_2}$ in which case the model does not have vacuum instability problem.

However there is
 the possibility of a new class of RG  solutions for unflavoured leptogenesis with much heavier triplet masses, e.g $M_{\Delta_2}\sim 10^{14}$ GeV and $M_{\Delta_1}=M_{15_{H1}}\sim M_U$, which does have the vacuum instability in the absence of any non-standard scalar field near $\sim {\cal O} 10^{10}$ GeV \cite{Espinosa:2012} or near the electroweak scale \cite{Lebedev,spc-NPB:2018} in the TTM.    

We note that the present SU(5) model  with  fine tuning 
in the GUT scale Higgs potential can remove such  negativity of the quartic coupling for all values of $|\phi| \ge 5\times 10^9 $ GeV even if $\Delta_i(i=1,2)$ threshold effects in eq.(\ref{eq:thlam}) are negligible. This advantage occurs in this GUT framework that permits the scalar singlet $S_{24}\subset {24}_H$ to be at the desired intermediate scale.
 The Higgs portal quartic coupling of this singlet with SM Higgs $\phi \subset {5}_H$ originates from the term
\bea
V_q&=&\lambda_{S}{24}_H^4+\lambda_{(\phi,S)}{24}_H^2{5}_H^{\dagger}{5}_H \nonumber\\  
  &\supset& \lambda_SS_{24}^4+\lambda_{(\phi,S)}\phi^{\dagger}\phi S_{24}^2. \label{eq:lambhs}     
\eea
Even though $\langle S_{24}\rangle \sim M_U$, it is possible to make the SM scalar singlet mass lighter $M_{S_{24}}\simeq (10^{8}-10^9)$ GeV and this gives rise to threshold effect \cite{Espinosa:2012}
\beq
\Delta {\lambda_{\phi}}^{(S)}=\frac {\lambda_{(\phi,S)}^2}{\lambda_{S}}\Theta(|\phi|-M_{S_{24}}), \label{eq:ths}
\eeq
where $\lambda_{S}=$  the singlet Higgs self coupling originating from 
$\lambda_{24}{24}_H^4\supset \lambda_{S}S_{24}^4$. 
The  solution to vacuum stability issue then proceeds by RG running the SM Higgs quartic coupling and taking into account this threshold effect at the field value $|\phi|\simeq M_{S_{24}}\sim {\cal O} (10^8-10^9)$ GeV in a manner exactly similar to our earlier work \cite{pcns:2020} which we do not repeat here \cite{cps:2019}.
Thus, irrespective of the heavier mass scales $M_{\Delta_1}\gg M_{\Delta_2}\gg M_{S_{24}}$, the presence of the intermediate mass scalar singlet $S_{24}$ at $M_{S_{24}}\sim (10^8- 10^9)$ GeV ensures stability of the standard Higgs potential in this SU(5) theory.
It is interesting to note that such a vacuum stability is maintained in the presence of both lighter ( under class (ii)) and heavier triplet fermion DM  mass
(under class (i)) solutions contained in the mass range $M_{\Sigma}\sim (500-3000)$ GeV. In what follows, we show how this solution to vacuum stability reduces the scalar singlet DM mass prediction.\\    
\section{Fermion triplet plus scalar singlet dark matter under indirect search constraints}\label{sec:fxidm}

As discussed above the DM annihilation cross section  $\Sigma^+\Sigma^-\to W^+W^-$ in the perturbative estimation \cite{Ma-Suematsu:2008} shows a peak that is capable of accounting for the cosmologically observed  relic density $(\Omega h^2)_{Obs}=0.1172-0.1224$ if the thermally produced triplet fermion DM mass $M_{\Sigma^0}\simeq 2.4$ TeV. However a number of indirect experimental search constraints discussed  
 in Sec.\ref{sec:sigind} forbid the $\Sigma$ mass at this resonant value. 
 For example, this dominant thermal DM mass $M_{\Sigma}=2.4$ TeV does not solve the cosmic ray (CR) lepton puzzle \cite{Hryzcuk:2014} or produces excess of $\gamma$ ray towards nearby galaxy clusters   \cite{Han,Hektor,HESS,MAGIC,Ajello,Dugger,Zimmer,Huang:12}. The resonance DM mass $M_{\Sigma}=2.4$ TeV can be ruled out in this search channel while all other masses  are allowed in the dwarf spheroids channel \cite{Hryzcuk:2014}. All mass values  $M_{\Sigma} \ge 2$ TeV are ruled out by the HESS data \cite{HESS}. 
Whereas the data   from WMAP-5\cite{Komatsu}  have ruled out the mass range $M_{\Sigma}=2.3-2.4$ TeV, the mass range $M_{\Sigma}=2.25-2.46$ TeV has been ruled out by WMAP-9\cite{Hinshaw}, and the combined search of WMAP-9 and ACT \cite{Sievers,Story} has also ruled out the mass range $M_{\Sigma}=2.18-2.5$ TeV. 
Similarly  the peak value of DM mass $M_{\Sigma} \ge 2.74$ TeV or heavier, determined through 
non-perturbative Sommerfeld enhancement \cite{Cirelli:2006,Frig-Ham:2010},  appears to be  ruled out as it predicts annihilation cross section exceeding the upper limit set  by the combined analysis of Fermi/LAT \cite{fermi} and MAGIC \cite{MAGIC}. Consequently, these indirect search experiments constraining $M_{\Sigma}$ to be significantly less than the 
resonance   value $2.4$ TeV (or less than the non-perturbatively determined peak value $ 2.74-3$ TeV ) deny the thermally produced triplet fermion $\Sigma^0$ to be the dominant WIMP DM of the Universe.\\     
On the other hand, we have found that  every $\Sigma-$ mass value in the range $M_{\Sigma}={\cal O}(500-3000)$ GeV is capable of producing SU(5) unification of the TTM with precision coupling unification and prediction of the same sets of heavy 
scalar triplet  masses matching the neutrino oscillation data, and leading to successful  unflavoured or partially flavoured leptogenesis. Each of the $\Sigma$   masses in the range $M_{\Sigma}={\cal O}(500-3000)$ GeV also predicts spin-independent elastic $\Sigma-p$ scattering cross sections in concordance with direct detection bounds \cite{Aprile:2017,Aprile:2018,Hisano:2015}. We have further pointed out  that all classes of solutions in  our unification model can be consistent with vacuum stability of the SM scalar potential caused
by threshold effect due to an intermediate mass Higgs scalar singlet $S_{24}\subset {24}_H$ of SU(5). \\
Thus, barring  indirect experimental search constraints, if collider experiments in future finally succeed in detecting 
the heavier triplet fermion DM with perturbatively (non-perturbatively) predicted resonance mass $M_{\Sigma}=2.4$ TeV ( $M_{\Sigma}\ge 2.74$ TeV or still heavier), our  
 unification solutions  of the type given in Table \ref{tab:RGEsol2}, Fig. 
\ref{fig:invHSIG}, and Fig. \ref{fig:taupHSig}, belonging to class (i) category, are consistent with such $\Sigma$ as the dominant thermal DM of the Universe accounting for the entire value of the observed cosmological relic density \cite{Planck15,wmap}. For this class  of heavier triplet mass solutions in our unification model, there is no need to invoke the presence of a scalar singlet DM. \\

  But  if the future collider searches fail to detect such resonant $M_{\Sigma}=2.4$ TeV ,or heavier, and  indirect experimental search constraints which forbid the resonant masses $M_{\Sigma}\simeq 2.4$ TeV (or corresponding non-perturbative values $M_{\Sigma} \simeq 2.74-3$ TeV)  are accepted, it is difficult to account for the
entire observed value of relic density with such lighter
$M_{\Sigma}$ values although they ensure TTM unification through SU(5). For such lighter $M_{\Sigma} < 2.0$
TeV values as sub-dominant  dark matter, we propose to justify our unification solutions of the type presented in Table \ref{tab:RGEsol}, Fig. \ref{fig:invf}, Fig. \ref{fig:gi}, and Fig. \ref{fig:taup5} also from dark matter point of view. In such other class of unification solutions corresponding to lighter $M_{\Sigma} <  2.0$ TeV values, we introduce a  SU(5)-singlet scalar  $\xi(1,0,1)$  as shown in Table \ref{tab:su5z2} without affecting  gauge coupling unification, neutrino mass, and leptogenesis results. We hypothesize the WIMP DM to comprise of two components: the neutral component of fermion triplet $\Sigma(3,0,1)$ of mass $M_{\Sigma}\sim {\cal O}(500-2000)$ GeV and the scalar singlet $\xi(1,0,1)$ whose mass is to be determined from relic density and direct detection masss bounds.  The  assignment of global discrete symmetry $Z_2(\xi)=-1$ as shown in Table \ref{tab:su5z2} stabilises it as a scalar singlet component of WIMP DM \cite{GAMBIT,pcns:2020}.\\

Similar two-component WIMP DM problem comprising of fermion triplet and scalar singlet has been addressed earlier \cite{Ma-Suematsu:2008} and very recently \cite{cps:2019} where the  DM was also constrained to satisfy vacuum stability  by ensuring the SM Higgs quartic coupling to run positive till $|\phi|\simeq M_{pl}=1.22\times 10^{19}$ GeV through correction derived from DM portal. In the presence of $\xi$ the SM Higgs potential is changed to
\begin{equation}
V=-\mu_{\phi}^2 \phi^\dagger \phi +\mu_\xi^2 \xi^\dagger \xi  + \lambda_\phi (\phi^\dagger \phi)^2 +\lambda_\xi (\xi^\dagger \xi)^2 +
2 \lambda_{\phi \xi}  (\phi^\dagger \phi)(\xi^\dagger \xi). \label{eq:dmpot}
\end{equation}
leading to respective masses of standard Higgs ($=m_{\phi}$) and scalar DM ($=m_{\xi}$)
 \begin{eqnarray}
&& m_{\xi}^2 = 2(\mu_\xi^2 +\lambda_{\phi \xi}^2 v^2), \nonumber\\
&& m_\phi^2 =2 \mu_{\phi}^2= 2 \lambda_\phi v^2.   \label{eq:mxiphi}  
\end{eqnarray}
But unlike \cite{cps:2019,pcns:2020}, there are several other sources which can ensure  vacuum stability in this model. Out of these the intermediate scale Higgs scalar singlet threshold effect of eq.(\ref{eq:ths}), as discussed above in Sec.\ref{sec:vstab}, is capable of ensuring vacuum stability for all classes of solutions including the dominant heavy triplet fermionic DM case ($M_{\Sigma}=2.4$ TeV, or $M_{\Sigma}\ge 2.74$ TeV).

This positive definite threshold correction is sufficient to ensure vacuum stability till the Planck scale even if threshold corrections due to two heavy scalar triplets $\Delta_1$ and $\Delta_2$ are neglected. The triplet fermion DM $\Sigma$ does not contribute to renormalisable corrections to SM Higgs potential or its vacuum stability. 
 In the  class of our unification solutions with lower $M_{\Sigma}$ values, the added scalar singlet DM $\xi$ is now constrained by the part of observed relic density \cite{Planck15,wmap} missed by the non-resonant $M_{\Sigma}$  and the direct detection mass bounds \cite{Akerib:2016,Aprile:2017,Aprile:2018,Cui:2017}.
As  the vacuum stability constraint \cite{pcns:2020} is resolved by the intermediate scalar threshold effect, we discuss below how lower masses of the scalar DM $\xi$ are predicted in this model.
\subsubsection{\textbf{Relic density constraint on scalar dark matter}}\label{sec:relxi}
A cosmologically disadvantageous point about the lighter $M_{\Sigma}$ solutions is that the non-perturbative 
Sommerfeld enhancement is not effective at such masses. Also the perturbative estimation of annihilation  cross section leads to the thermal relic abundance, ${(\Omega h^2)}_{\Sigma}$ significantly less than the experimentally observed value determined 
by Planck\cite{Planck15} and WMAP\cite{wmap}. This deficit is
compensated  through the intervention  of the scalar singlet DM
component $\xi$. Since the scalar singlet can not have any
renormalisable interaction with  $\Sigma$, we can 
estimate the relic density for fermionic triplet ($=(\Omega h^2)_{\Sigma}$)
and the scalar
 singlet $(=(\Omega h^2)_{\xi})$ separately, and  add them up to get the total 
relic density 
\begin{equation}
 (\Omega h^2)_{Obs}=(\Omega h^2)_{\Sigma} + (\Omega h^2)_{\xi} . \label{eq:fsrelic}
\end{equation}
where 
\beq
(\Omega h^2)_{Obs}=0.1172-0.1224, \label{eq:omobs}
\eeq
  is the observed value of cosmological DM relic density \cite{Planck15,wmap}.
For notational convenience, we also equivalently define the normalised relic density $R_{\xi}(R_{\Sigma})$ for thermal DM component $\xi (\Sigma)$,
\bea
R_{\xi}&=& \frac{(\Omega h^2)_{\xi}}{(\Omega h^2)_{Obs}}, \nonumber\\
R_{\Sigma}&=& \frac{(\Omega h^2)_{\Sigma}}{(\Omega h^2)_{Obs}},\nonumber\\  
R_{\Sigma}&+&R_{\xi}=1, \label{eq:omrs}
\eea
In the estimation of total relic density through  eq.(\ref{eq:fsrelic}), essentially there are three parameters $M_{\Sigma}, m_{\xi}$ and the $\xi-\phi$ Higgs portal coupling $\lambda_{\phi \xi}$. 
To estimate the relic abundance of the neutral component $\Sigma^0$ that  acts as sub-dominant thermal dark matter, we use eq.(\ref{eq:omh2}) and the iterative solution from eq.(\ref{eq:xf}) taking  into account the annihilations and co-annihilations as explained in Sec. \ref{sec:fdm}. 
    
Since in the lower mass region of fermion triplet DM, $M_{\Sigma}={\cal O}(500-1000)$ GeV, the predicted value of $(\Omega h^2)_{\Sigma}$ is found to be much less than the experimentally observed value  \cite{Planck15,wmap} given in eq.(\ref{eq:omobs}), the total relic density is dominated by the scalar DM contribution in this $M_{\Sigma}$ region leading to the constraint $(\Omega h^2)_{Obs}\simeq (\Omega h^2)_{\xi}$.

 Approximate solution of the corresponding Boltzmann equation for the singlet scalar DM component $\xi$ also  gives the expression for relic density 
\begin{equation}
(\Omega_{\rm DM} h^2)_{\xi}= \frac{1.07\times10^9 x_F^{\xi}}{\sqrt{g_\ast} M_{ pl}\langle \sigma^{\xi} v \rangle} \label{eq:srelic}
\end{equation}
where $x_F^{\xi}=M_{\xi}/T_F$,  $T_F=$ freezeout temperature, $g_\ast=$  effective number of massless degrees of freedom and  $M_{pl}=1.22\times10^{19}$ GeV. Now  $x_F^{\xi}$ can be computed by iteratively solving the equation
\begin{equation}
x_F^{\xi}=\ln \left(\frac{M_\xi}{2\pi^3}\sqrt{\frac{45 M_{pl}^2}{8 g_\ast x_F^{\xi}}} \langle\sigma^{\xi} v \rangle \right) .\label{eq:xfxi}
\end{equation}
In  eq.(\ref{eq:srelic}) and eq.(\ref{eq:xfxi}), the only particle physics
input is  the thermally averaged annihilation
cross section.
The total annihilation cross section is obtained by summing over all the annihilation channels of the singlet DM which are $\xi\xi \rightarrow F\bar{F},W^+W^-,ZZ,hh$
where $F$ is symbolically used for all the fermions. Using the expression of total annihilation 
cross section\cite{McDonald:1993ex,Guo:2010hq} in eq.(\ref{eq:xfxi}) at first we compute
$x_F^{\xi}$ which is then used in eq.(\ref{eq:srelic}) to yield the relic density contribution due to $\xi$.
As already noted above, in the estimation of the total relic density  
there are three free parameters: mass of the fermion triplet $(M_\Sigma)$, mass of the scalar singlet $(m_\xi)$ and the Higgs portal coupling $(\lambda_{\phi\xi})$.
We have examined the  outcome of this combined DM hypothesis, $\Sigma \oplus \xi$, upon the following values of $M_{\Sigma}$ allowed by indirect search constraints: {\bf (a) $M_{\Sigma}\simeq (500-1000)$\,GeV:} where $R_{\Sigma}\simeq 0, R_{\xi}\simeq 1$ as explained below, and {\bf (b) $M_{\Sigma} \simeq 1.5$ TeV:} where $R_{\Sigma}\simeq R_{\xi}=0.5$.\\


\par\noindent{\bf (a) $M_{\Sigma}\simeq (500-1000)$\,GeV:}\\
\bea
(\Omega h^2)_{\Sigma}&\simeq&0,  \nonumber\\
R_{\Sigma}&\simeq& 0,      \nonumber\\        
R_{\xi}&\simeq& 1.0.       \label{eq:rxi1}    
\eea
As the quantity $(\Omega h^2)_{\Sigma}\simeq 0$ for values of $M_{\Sigma} \simeq (500-1000)$ GeV, the scalar DM contribution to the relic density is required to meet the experimental constraint due to WMAP \cite{wmap} and Planck \cite{Planck15} measurements: 
\beq
 0.1172<(\Omega h^2)_{\xi} <0.1224. \label{eq:xirel}
\eeq
It is worth while to mention that we have varied
$\lambda_{\phi\xi}$ and $m_\xi$ over a wide
range of values to get their first round of constraint from the relic density bound given above. 
Two free parameters involved in this computation are mass of the 
 DM particle $m_\xi$ and the Higgs portal coupling $\lambda_{\phi \xi}$. The relic density has been estimated for a wide range of values of the scalar
DM  mass ranging from few GeVs to few TeVs while the coupling $\lambda_{\phi \xi}$ is also varied simultaneously in the range $(10^{-4}-1)$.
The parameters $(m_\xi,\lambda_{\phi\xi})$ are constrained by using the bound on the relic density  reported by WMAP \cite{wmap} and Planck \cite{Planck15} mentioned above in eq.(\ref{eq:xirel}). In Fig.\ref{fig:reldir} through the yellow curve,
we show only those combinations of $\lambda_{\phi \xi}$  and $m_\xi$ which are capable of producing the entire value of the experimentally observed relic density.
This yellow curve in Fig. \ref{fig:reldir} has been labelled as  $(R_{\xi}=1,R_{\Sigma}=0)$.\\
\par\noindent{\bf (b) $M_{\Sigma}=1500$\,GeV:}\\
Unlike the case (a) discussed above where $\Sigma$ has almost negligible contribution to relic density, the fermionic sub-dominant DM has very significant
 contribution \cite{Hisano:2007} to the experimentally observed value of relic density \cite{Planck15,wmap}  at $M_{\Sigma}\simeq 1.5$ TeV. 
At first we estimate $(\Omega h^2)_{\Sigma}$ at $M_{\Sigma}=1500$ GeV using the procedure outlines in Sec.\ref{sec:fdm} and  \cite{Hisano:2007} leading to
\bea
(\Omega h^2)_{\Sigma}&\simeq&\frac{1}{2}(\Omega h^2)_{Obs},  \nonumber\\
R_{\Sigma}&\simeq& \frac{1}{2}.     \label{eq:relsigb}
\eea
Then using eq.(\ref{eq:fsrelic}) and eq.(\ref{eq:omrs}) gives
\bea
(\Omega h^2)_{\xi}&\simeq&\frac{1}{2}(\Omega h^2)_{Obs},  \nonumber\\
R_{\xi}&\simeq& \frac{1}{2}.      \label{eq:relxib}
\eea
Following the similar procedure explained in the case (a) above, we have plotted
$\lambda_{\phi,\xi}$ against $m_{\xi}$ while keeping the value $R_{\xi}\simeq 0.5$
throughout. In Fig. \ref{fig:reldir} these results have been presented by the blue curve labelled as $(R_{\xi}=R_{\Sigma}=1/2)$.
\subsubsection{\textbf{Bounds from direct detection experiments}}
We have explicitly shown in Sec. \ref{sec:dircoll} and  Fig. \ref{fig:xesig}
that for all values of fermionic triplet masses $M_{\Sigma}=(500-3000)$ GeV, the 
respective spin-independent (SI) $\Sigma-p$ elastic cross sections are well below the direct detection bounds due to XENON1T collaboration\cite{Aprile:2017,Aprile:2018}.  
With the set of values of $(\lambda_{\phi \xi},m_\xi)$
already restricted by relic density bound, we proceed further to constrain them by  direct detection bound applied to $\xi-$ N elastic scattering. For this two component dark matter, the constraint 
relation appears as \cite{Ma-Suematsu:2008,Cao:2007fy}
\begin{equation}
\frac{\epsilon_\Sigma}{M_\Sigma} \sigma_{\Sigma N} + \frac{\epsilon_\xi}{m_\xi} \sigma_{\xi N} < \frac{\sigma_0}{m_0}, \label{eq:ineq} 
\end{equation}
where the symbols have the following meanings: $\sigma_0,m_0$ are SI DM-nucleon scattering cross section and mass of the dark matter, respectively, for single component
dark matter scenario. For the two component dark matter scenario under consideration $\sigma_{\Sigma N}(\sigma_{\xi N})$ is the SI scattering cross section
of $\Sigma(\xi)$ with detector nucleon and $M_{\Sigma}(m_\xi)$ is its mass. The factor $\epsilon_i$ designates the fraction of  density of the $i$th dark matter particle in a
certain model: $\epsilon_i=\rho_i/\rho_0$ which can also be expressed in terms of thermally averaged annihilation cross sections as 
\begin{eqnarray}
&&\epsilon_{\Sigma} =\frac{\langle \sigma v \rangle_\xi}{ \langle \sigma v \rangle_\xi +\langle \sigma v \rangle_\Sigma } \nonumber\\
&&\epsilon_\xi=\frac{\langle \sigma v \rangle_\Sigma}{ \langle \sigma v \rangle_\xi +\langle \sigma v \rangle_\Sigma } ~.\label{eq:compdm}
\end{eqnarray}
Noting that $\epsilon_{\Sigma}\le 1$ and $\epsilon_{\xi}\le 1$, in the parameter space permitted by indirect experimental searches restricting
$M_{\Sigma}\simeq {\cal O}(500-2000)$ GeV, it is easy to verify that the inequality   given by eq.(\ref{eq:ineq}) is satisfied.\\
 We get exclusion plots of DM-nucleon scattering cross section and DM mass from different direct
detection experiments \cite{Akerib:2016,Aprile:2017,Aprile:2018,Cui:2017}. The spin independent scattering cross section of singlet DM on nucleon is given by\cite{Cline:2013gha}
\begin{equation}
\sigma^{\rm SI}=\frac{4 f_R^2\lambda_{\phi \xi}^2 \mu_R^2 m_N^2}{\pi m_\xi^2 m_h^4} ~~({\rm  cm}^2) \label{sigsi}
\end{equation}
where $m_h=$ mass of the SM Higgs ($\sim 125$ GeV), $m_N=$ nucleon mass $\sim 939$ MeV, $\mu_R=(m_\xi m_N)/(m_\xi+m_N)=$  reduced 
DM-nucleon mass and the factor $f_R \sim 0.3$. Using eq.(\ref{sigsi}) the exclusion plots of $\sigma-m_\xi$ plane can be easily brought to $\lambda_{\phi \xi}-m_\xi$
plane. We superimpose the $\lambda_{\phi \xi}-m_\xi$ plots for different experiments  \cite{Akerib:2016,Aprile:2017,Aprile:2018,Cui:2017} on the plot of allowed parameter space constrained by relic density bound.

\begin{figure}[h!]
\begin{center}
\includegraphics[scale=0.6,angle=0]{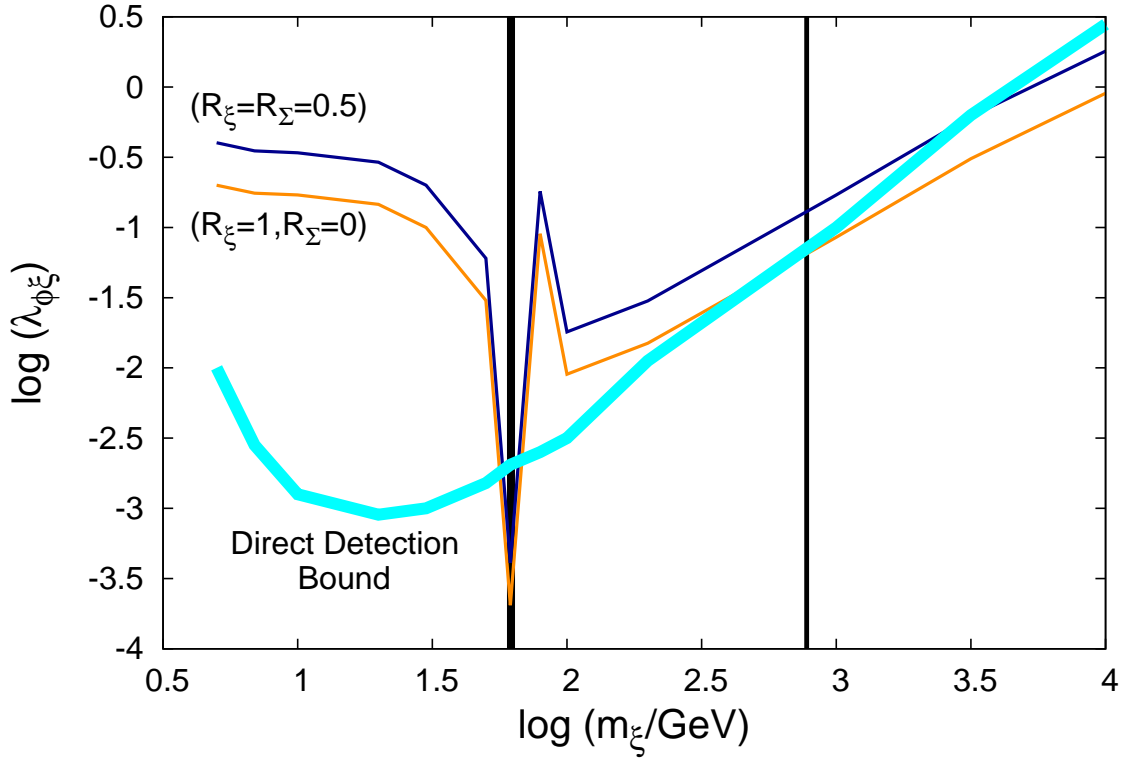}
\end{center}
\caption{Scalar singlet DM mass predictions from constraints due to indirect seaches on $M_{\Sigma}$, observed  relic density \cite{Planck15,wmap},  and direct detection cross section bounds  (green curve): (i) Yellow curve,$(R_{\xi}\simeq 1,R_{\Sigma}\simeq =0)$,
$500 {\rm \, GeV}\le M_{\Sigma} < 1000 {\rm GeV}$; (ii) Blue curve, $(R_{\xi}=R_{\Sigma}\simeq 0.5)$, $M_{\Sigma}=1.5$ TeV. Common value of $\xi$ mass as low as $m_{\xi}=62$ GeV for both the yellow and the blue curves is indicated by the first vertical line. The second vertical line indicates prediction of  $m_{\xi}=790$ GeV when $R_{\xi}=1$ and $\lambda_{\phi\xi}\simeq 0.08$, The crossing of the blue line with the green band indicates prediction of a heavier $m_{\xi}\simeq 3.1$ TeV with
$\lambda_{\phi\xi}\simeq 0.56$ when $R_{\xi}\simeq 0.5$.}   
\label{fig:reldir}
\end{figure}

Thus the parameter space $(\lambda_{\phi\xi}~vs~m_\xi)$ constrained by both the relic density  and the direct detection experimental bounds can be obtained from Fig.\ref{fig:reldir}.
Using complete dominance of $\xi$ in the case (a) of the lighter values of $500 < M_{\Sigma}/{\rm GeV} \le 1000$ and using $R_{\xi}\simeq 1$ we have  different sets of values of $(\lambda_{\phi\xi}, m_{\xi})$ while matching the observed relic density due to Planck \cite{Planck15}
and WMAP \cite{wmap}: $(\Omega h^2)_{\xi}\simeq (\Omega h^2)_{Obs} =0.1172-0.1224$  through eq.(\ref{eq:rxi1}).
This relic density matching  is shown by the yellow curve in Fig.\ref{fig:reldir} which has been also labelled as $(R_{\xi}=1,R_{\Sigma}=0)$.
 The green band represents  
 exclusion plots obtained from dark matter direct detection experiments, LUX-2016\cite{Akerib:2016}, XENON1T \cite{Aprile:2017,Aprile:2018} and 
 Panda-XII(2017) \cite{Cui:2017}. All points  below this green band are allowed by direct detection experiments.\\

In Fig.\ref{fig:reldir} the points on the yellow curve or the blue curve which are also below the green band  are allowed by both the relic density
as well as the bounds on the DM-nucleon annihilation cross section as reported by the direct detection experiments. From the Fig.\ref{fig:reldir} we find that for both the yellow and the blue curves, the scalar singlet DM with masses  $m_{\xi}=10-60$ GeV and $65 \le m_{\xi}/{\rm GeV}$  are ruled out by the combined constraints of relic density and direct detection  limits . For small DM-Higgs portal coupling $\lambda_{(\phi,\xi)}\simeq (10^{-4}-10^{-3})$, the lowest predicted value of the scalar DM mass is $m_{\xi}\simeq 62$ GeV which is shown by the first vertical solid line in 
Fig. \ref{fig:reldir}. \\
We further note that in the case of the yellow curve all scalar DM masses  $m_{\xi} \ge 790$ GeV satisfy the combined constraints due to relic density and direct detection  limits. Out of this latter set of solutions, the lowest allowed mass in this range is $m_{\xi}\simeq 790$ GeV having the  DM-Higgs portal coupling $\lambda_{(\phi,\xi)}\simeq 0.08$. These 
predictions are presented in  Table \ref{tab:ximass}.\\
\begin{table}[h!]
\caption{Prediction of  singlet scalar component  dark matter mass from constraints due to indirect searches, observed DM relic density, and elastic cross section bounds from direct detection experiments. For renormalisation group running, initial values of SM gauge couplings
$g_i (i=Y,2L,3C)$ and the top-quark
Yukawa coupling $h_t$  have been fixed at top quark mass
  $\mu=m_{top}=173.34$ GeV using PDG data \cite{PDG:2014}. The SM Higgs quartic coupling has been fixed at $\lambda_{\phi}=0.129$ corresponding to physical Higgs mass $m_h=125$ GeV \cite{cps:2019,pcns:2020}.}
\begin{tabular}{|c|c|c|c|c|c|c|c|c|c|}  
 \hline
  $M_{\Sigma}$(GeV)&$R_{\xi} $&$R_{\Sigma}$&$m_\xi$ (GeV) & $\lambda_{\phi \xi} $ &  $\lambda_{\xi} $ & $g_{1Y}$ & $g_{2L}$ &  $g_{3C}$ &  $h_{t}$   \\ 
 & & & & & & & & & \\ \hline
$500-1000$&$1.0$&$0.0$&$62$&$1.7\times 10^{-4}-1.6\times 10^{-3}$& $0.25$&$0.35$&$0.64$&$1.16$&$0.96$\\ 
\hline
$500-1000$&$1.0$&$0.0$&$790$&$0.08$&$0.19$&$0.35$&$0.64$&$1.16$&$0.96$\\ 
\hline
$1500$&$0.5$&$0.5$&$62$&$1.7\times 10^{-4}-1.6\times 10^{-3}$&$0.25$&$0.35$&$0.64$&$1.16$&$0.96$\\
\hline
$1500$&$0.5$&$0.5$&$3100$&$0.56$&$0.27$&$0.35$&$0.64$&$1.16$&$0.96$\\
\hline
\hline
\end{tabular} 
\label{tab:ximass}
\end{table}
 
In the case of the blue curve representing nearly $50\%$ of observed relic density due to $\xi$ ( the other $\simeq 50\%$ being due to $\Sigma$ ), while one of the possible $m_{\xi}$ predictions is similar to the case (a) with $m_{\xi}\simeq 62$ GeV, the heavier mass solution
has been shifted to $m_{\xi}\ge 3.1$ TeV as indicated by the crossing point of this curve with the green band in Fig. \ref{fig:reldir} . In fact  the heavier $\xi$ mass prediction appears to depend upon $R_{\xi}$. Consistency of all $M_{\Sigma}$ values with
direct detection bounds have been already discussed above. 

\section{Grand unification advantage in dark matter predictions}\label{sec:GA}
In the TTM  investigations without any GUT origin, the most convenient DM embedding was a scalar singlet DM of mass $\simeq 1.3$ GeV \cite{pcns:2020}. In addition to fulfilling the constraints due to relic density  and direct detection cross section bounds,  the higher value of the scalar singlet DM mass prediction was a consequence of vacuum stability constraint of the SM Higgs potential that existed in the TTM without this scalar singlet. When TTM is unified 
into SU(5), we find the necessity of fermionic triplet  DM $\Sigma (3, 0, 1)$ in addition to an intermediate mass colour octet fermion $C_F(1,0,8)$ for precision gauge coupling unification. Further,
the unified model predicts  a SM-singlet Higgs scalar $S_{24}$ originating from ${24}_H$ of SU(5) which can be easily made to have an intermediate mass  $M_S \simeq (10^8-10^9)$ GeV. As such, the corresponding threshold effect due to this scalar ensures vacuum stability for all classes of solutions for leptogenesis and type-II seesaw ansatz for
neutrino masses in the TTM unified SU(5). This intermediate mass Higgs scalar guarantees vacuum stability with all allowed values of fermionic triplet DM masses
discussed above. However, the class of lighter mass fermion  triplet DM predicted by coupling  unification being inadequate for relic density,
requires the introduction of a scalar singlet DM $\xi$  through $SU(5)\times Z_2$. Thus the GUT theory providing an alternative solution for vacuum stability through the intermediate mass Higgs scalar, now permits much lower mass solutions for the scalar DM component $m_{\xi}\simeq 62$ GeV and $m_{\xi}\simeq 790$ GeV which were ruled out by the TTM without grand unification \cite{pcns:2020}.\\

For heavier $M_{\Sigma}\simeq 1500$ GeV  allowed by indirect search constraints, the GUT embedding of TTM also predicts the lighter scalar DM mass almost identical to  $m_{\xi}\simeq 62$ GeV, although a  heavier mass $m_{\xi}\simeq 3.1$ TeV is also predicted with $\lambda_{\phi,\xi}\simeq 0.56$. These solutions indicate that
the heavier $m_{\xi}$ prediction is a consequence of the value of $M_{\Sigma}$ which controls the value of $R_{\xi}$.
Details of such investigations would be reported elsewhere \cite{mkp-rs:prep}.

 The radiative stability of the Higgs mass which has been shown to be possible in  TTM even without its GUT embedding  
\cite{pcns:2020}, is also consistently realised in this SU(5) model  which permits   \cite{del Aguilla:1981,RNM-gs:1983}   fine-tuning of parameters  in the presence of radiative corrections.


\section{Summary and outlook}\label{sec:sum}
This grand unification program is a natural follow up of a  recent investigation \cite{pcns:2020} where it has been shown that the two heavy scalar triplet model (TTM) \cite{Ma-Us:1998,Sierra:2014tqa} can successfully explain the current neutrino oscillation data with hierarchical neutrino masses in concordance with  cosmological bounds \cite{Planck15,Sunny:2018} while predicting the observed baryon asymmetry of the Universe through unflavoured or partially flavoured ($\equiv \tau-$ flavoured) leptogenesis without requiring any right-handed neutrino (RHN). Noting that: (i) unlike SO(10) or $E_6$, the SU(5) grand unified theory does not have RHNs 
in its fundamental fermion representations suiting to the basic requirement  of the two-triplet model (TTM), and (ii) SU(5) needs  smaller representations ${15}_{H1}\oplus {15}_{H2}$ compared to much larger representations ${126}_{H1}\oplus {126}_{H2} \subset $ SO(10) (or ${351}^{\prime}_{H1}\oplus {351}^{\prime}_{H2}\subset E_6$),  for the first time through this work we have delineated the outstanding position of SU(5) compared to higher rank GUTs to achieve TTM unification.\\
  
In addition to a colour octet fermion 
$C_F(1,0,8)$ of mass $M_{C_F} \ge 10^{8}$ GeV,  the unification completion is  found to  predict the well known fermionic triplet  dark matter $\Sigma(3,0,1)$ in the mass range $M_{\Sigma}\sim {\cal O}(500-3000)$ GeV  both of which originate from the non-standard fermionic representation ${24}_F\subset$ SU(5). Every  input value of $\Sigma$ mass in the investigated range $M_{\Sigma} \simeq {\cal O}(500-3000)$ GeV with suitable combinations of masses  $M_{C_F}, M_{\Delta_1}=M_{15H1}, M_{\Delta_2}$, and  $M_U$ is found to achieve successful unification of all the six different  solutions of the ununified TTM of Table \ref{tab:masspara}. On the basis of WIMP DM paradigm, the unification solutions achieved as a function of $M_{\Sigma}$ have been categorised into two classes. Under class (i) solutions applicable for heavier $M_{\Sigma}> 2$ TeV,  resonance mass of dominant fermionic DM accounts for the entire value of the observed  cosmological relic density. Examples of class (i)  solutions corresponding to a chosen resonance mass value of $M_{\Sigma}=2.4$ TeV are presented  numerically in Table \ref{tab:RGEsol2} 
where all the six ununified TTM solutions derived in \cite{pcns:2020} have been successfully unified through SU(5). Precision gauge coupling unification and proton lifetime estimations as a function of $V_{eff}$ have been presented in Fig. \ref{fig:invHSIG} and Fig. \ref{fig:taupHSig} under this class (i) solution corresponding to $M_{\Sigma}=2.4$ TeV.  However, such resonance mass values $M_{\Sigma}\simeq 2.4$ TeV, or heavier, appear to have been challenged by a number of indirect search experiments
 as  summarised in Sec.\ref{sec:sigind} and Sec. \ref{sec:fxidm}. Then, barring such indirect search constraints, future collider searches may decide on the existence of such a dominant fermionic DM which would go in favour of our class (i) solutions.\\ 
A substantial region of our successful precision unification results consists of class (ii)  solutions  with lighter sub-dominant DM $\Sigma$ masses $M_{\Sigma}\simeq {\cal O}(500-2000)$ GeV which are not constrained by indirect experimental search results. As these lighter masses can not account for the entire value of the observed cosmological relic density, the resulting deficit  is circumvented by the introduction of a scalar singlet DM $\xi$. Consistent with the combined constraints of observed relic density and direct detection cross section bounds,
 we have carried out the scalar singlet mass ($m_{\xi}$) predictions in relation to $M_{\Sigma}$ under indirect search constraints. In the region $M_{\Sigma}=(500-1000)$ GeV where $\Sigma$ has negligible relic density contribution, this scalar singlet $\xi$  mass is predicted to be $m_{\xi}\simeq 62$ GeV  for a smaller Higgs portal coupling $\lambda_{\phi\xi}=(10^{-4}-10^{-3})$ or $m_{\xi}=790$ GeV for more reasonable value of
 $\lambda_{\phi\xi}\simeq 0.08$. These lighter $\xi-$ DM masses, which were prevented in the ununified TTM \cite{pcns:2020} because of the latter's underlying vacuum stability constraint \cite{pcns:2020}, are now permitted through the SU(5) grand unification that provides an alternative solution for the SM vacuum stability. Under indirect search constraint  we have examined the case of 
$M_{\Sigma}=1500$ GeV which contributes nearly $50\%$ to the observed cosmological relic density ( the other $\simeq 50\%$ being due to $\xi$). Although the lighter $m_{\xi}$ and the correspondingly associated smaller values of $\lambda_{\phi\xi}$ remain almost identical as shown in Table \ref{tab:ximass}, the heavier mass prediction is shifted to $m_{\xi}\simeq 3.1$ TeV. Our investigations clearly indicate that the heavier
scalar singlet DM mass  prediction (other than $m_{\xi}=62$ GeV) can be controlled by $M_{\Sigma}$ while respecting indirect search constraints. Details of invesigation currently in progress in search of  lighter $m_{\xi}$ ($< 3.1$ TeV) predictions with reasonable values of $\lambda_{\phi \xi}$ will be reported elsewhere \cite{mkp-rs:prep}.\\

 An example of unification solutions under class (ii) category for $M_{\Sigma}=500$ GeV  is  presented  through Table \ref{tab:RGEsol}, Fig. \ref{fig:invf}, Fig. \ref{fig:gi}, and Fig. \ref{fig:taup5}. As in the case of class (i)  unification results  enumerated in Table \ref{tab:RGEsol2}, our class (ii)   solutions  numerically presented in Table \ref{tab:RGEsol} also successfully unify all the six different respective ununified TTM solutions of Table \ref{tab:masspara} and  \cite{pcns:2020}.\\ 
    
We have further shown a comparison through Fig. \ref{fig:xesig} elucidating how spin-independent $\Sigma-p$ elastic cross section predictions \cite{Hisano:2015} for both the lighter or heavier $\Sigma$ masses in the range $M_{\Sigma}\simeq {\cal O}(500-3000)$ GeV, belonging to both class (i) and class (ii) solutions, are consistent with direct detection cross section bounds determined experimentally \cite{Aprile:2017,Aprile:2018}.\\

The vacuum stability of the SM scalar potential in all types of solutions 
under class (i) or class (ii) is also
ensured within the purview of the present SU(5) unification framework through threshold effect of an intermediate mass scalar  ($M_{S_{24}}\simeq 10^8-10^9$ GeV) originating from the GUT breaking.
All the colour octet fermion masses needed to complete gauge coupling unification under both class (i) and class (ii)  solutions are noted to satisfy the
desired cosmological bound \cite{Arkani:2004}.\\

  Even though the present SU(5) model solutions have somewhat lower unification scales, $M_U\simeq {\cal O}(10^{15})$ GeV, they are associated with an unknown mixing parameter $V_{eff}$ in the proton decay width formula \cite{Perez:2019-1,Perez:2018-2,Perez-1} which prevents proton lifetime to be predictive for $p\to e^+\pi^0$ mode; rather observation of proton decay is expected to fix this model parameter. In this work we have discussed how  the current Hyper-Kamiokande limit
 partially constrains this parameter space. However, a substantial region of the parameter space with $V_{eff} < 1$  is noted to survive despite improved 
 accuracy in  proton lifetime measurements. Any smaller changes over the  predominantly one-loop RG determination of  unification scales presented here  would be  similarly consistent with Hyper-Kamiokande limit on the proton lifetime 
through correspondingly adjusted values of $V_{eff}$.  \\

In conclusion we find that the present SU(5) GUT framework is quite effective  for precision gauge coupling unification, fitting neutrino oscillation data,  explaining the observed cosmological baryon asymmetry  of the Universe through leptogenesis, and understanding the WIMP DM matter paradigm while ensuring vacuum stability of the SM scalar potential.\\

\section{ Acknowledgement}
M. K. P. acknowledges financial support through the research project SB/S2/HEP-011/2013 awarded by the  Department of Science and Technology, Government of India.
R. S. is financed under Ph.D. researh fellowship grant of Siksha 'O' Anusandhan, Deemed to be University.

\section{Appendix}

\subsection{ Lighter weak triplet and colour octet fermion masses}\label{sec:A1}
 For the present purpose, we utilize the most convenient  Yukawa Lagrangian including both normalizable as well as non-renormalizable (NR) terms \cite{Bajc-Nem-gs:2007,Bajc-gs:2007}
\bea
{\cal L}_{NR}&=&M_{F}Tr({24}_F^2) + Y_{24}Tr({24}_F^2{24}_H) \nonumber\\
 &+&\frac{1}{M_{NR}}[k_1Tr({24}_F^2)Tr({24}_H^2)+ k_2[Tr({24}_F{24}_H)]^2 \nonumber\\
&+&k_3Tr({24}_F^2{24}_H^2)+k_4Tr({24}_F{24}_H{24}_F{24}_H)]  \label{eq:GYUKA}
\eea
For the mass scale in the non-renormalizable term we use $M_{NR}=M_{String}\sim 10^{17}$(or $M_{NR}\simeq 10^{18}$ GeV$=$ reduced Planck scale). When  the Yukawa couplings are switched off, all component fermions in ${24}_F$ are near the GUT scale with degenerate mass $M_F$. When SU(5) is broken by the VEV
\beq
\langle S_{24H} \rangle =\frac{V_U}{\sqrt 30}diag(2,2,2,-3,-3) \label{eq:vevua}
\eeq 
with  $V_U\sim M_{GUT}$, all component fermions in ${24}_F$ are expected to split
in their masses \cite{Bajc-gs:2007}
\bea
M_{(LQ)_F}&=&M_{F}-\frac{Y_{24}V_U}{2\sqrt 30}+\frac{V_U^2}{M_{NR}}\left(k_1+\frac{
(13k_3-12k_4)}{60}\right),\nonumber\\
M_{S_F}&=&M_{F}-\frac{Y_{24}V_U}{\sqrt 30}+\frac{V_U^2}{M_{NR}}\left(k_1+\frac{7}{30}(k_3+k_4)\right),\nonumber\\
M_{\Sigma}&=&M_{F}-\frac{3Y_{24}V_U}{\sqrt 30}+\frac{V_U^2}{M_{NR}}\left(k_1+\frac{3}{10}(k_3+k_4)\right),\nonumber\\
M_{C_F}&=&M_F+\frac{2Y_{24}V_U}{\sqrt 30}+\frac{V_U^2}{M_{NR}}\left(k_1+\frac{2}{15}(k_3+k_4)\right) \label{eq:FMAA}
\eea
Our RG constraints on gauge coupling  unification need $M_{\Sigma} \sim 500-1000$ GeV, but $M_{C_F}\sim 10^8-10^9$ GeV compared to GUT scale values of $M_F\sim V_U \sim {\cal O}(10^{16})$ GeV. We discuss below how  with appropriate fine tuning of parameters in eq.(\ref{eq:FMAA}), these two desired masses can be made lighter than the GUT scale while keeping other component masses in ${24}_F$ super heavy near $\simeq M_U$.
For the lightest weak triplet fermion mass we use
\bea
M_{\Sigma}&\simeq& \frac{3V_U^2}{10M_{NR}}(k_3+k_4), \nonumber\\
  &\simeq& {\cal O}(500-1000)\,\,{\rm GeV}, \label{eq:sigma} 
\eea
Then using the first relation of eq.(\ref{eq:sigma}) in the third relation of eq.(\ref{eq:FMAA})  gives
\beq
M_{F}-\frac{3Y_{24}V_U}{\sqrt 30}+\frac{V_U^2}{M_{NR}}k_1\sim 0.\label{eq:conra}
\eeq
Then using  eq.(\ref{eq:FMAA}), eq.(\ref{eq:sigma}), and eq.(\ref{eq:conra}) gives 
\bea 
M_{C_F}&=&\frac{5Y_{24}V_U}{\sqrt 30}+\frac{4}{9}M_{\Sigma} \nonumber\\
 &\simeq& \frac{5Y_{24}V_U}{\sqrt 30} \label{eq:solMCFA},
\eea
where the second line follows from the fact that $M_{\Sigma}\ll M_{C_F}$ which is desired for our unification solutions discussed in Sec. \ref{sec:su5}. Now we get $M_{C_F}={\cal O}(10^8 -10^9)$ GeV for Yukawa coupling values $Y_{24}\simeq 10^{-7}-10^{-5}$. 
 Noting that $k_3+k_4 \simeq 0$ and utilising relations given under eq.(\ref{eq:sigma}), eq.(\ref{eq:conra}), and eq.(\ref{eq:solMCFA}) in the  corresponding expressions under eq.(\ref{eq:FMAA}), it is straight forward to show that all  component masses of ${24}_F$ become super heavy except $\Sigma$ and $C_F$ as noted above.\\


\end{document}